\documentclass[aps,prd,twocolumn,groupedaddress,showpacs,eqsecnum,floatfix]{revtex4-1}

\usepackage{amsmath,amssymb}
\usepackage{graphicx}
\usepackage{bm}
\usepackage{color}

\newcommand{\bmath}[1]{\mbox{\boldmath{$#1$}}}

\begin{document}

\title{Vacuum polarization on the brane}

\author{Cormac Breen}
\email{cormac.breen@dit.ie}

\affiliation{School of Mathematical Sciences,
Dublin Institute of Technology, Kevin Street, Dublin 8, Ireland.}

\author{Matthew Hewitt}
\email{app10mh@sheffield.ac.uk}

\affiliation{Consortium for Fundamental Physics, School of Mathematics and Statistics,
The University of Sheffield, Hicks Building, Hounsfield Road, Sheffield S3 7RH, United Kingdom.}

\author{Adrian C. Ottewill}
\email{adrian.ottewill@ucd.ie}

\affiliation{School of Mathematical Sciences and Complex \& Adaptive Systems Laboratory,
University College Dublin, Belfield, Dublin 4, Ireland.}

\author{Elizabeth Winstanley}
\email{e.winstanley@sheffield.ac.uk}

\affiliation{Consortium for Fundamental Physics, School of Mathematics and Statistics,
The University of Sheffield, Hicks Building, Hounsfield Road, Sheffield S3 7RH, United Kingdom.}

\date{\today}

\begin{abstract}
We compute the renormalized expectation value of the square of a massless, conformally coupled, quantum scalar field on the brane of a higher-dimensional
black hole.  Working in the AADD brane-world scenario, the extra dimensions are flat and we assume that the compactification radius is large compared with the size of the black hole.  The four-dimensional on-brane metric corresponds to a slice through a higher-dimensional
Schwarzschild-Tangherlini black hole geometry and depends on the number of bulk space-time dimensions.
The quantum scalar field is in a thermal state at the Hawking temperature.
An exact, closed-form expression is derived for the renormalized expectation value of the square of the quantum scalar field on the event horizon of the black hole. Outside the event horizon,
this renormalized expectation value is computed numerically.
The answer depends on the number of bulk space-time dimensions, with a magnitude which increases rapidly as the number of
bulk space-time dimensions increases.
\end{abstract}


\maketitle

\section{Introduction}
\label{sec:intro}

Higher-dimensional brane-world models \cite{ArkaniHamed:1998rs,Antoniadis:1998ig,ArkaniHamed:1998nn,Randall:1999ee,Randall:1999vf} have the consequence
that the energy scale of quantum gravity may be many orders of magnitude smaller than
the usual Planck scale $10^{19}$~GeV, and may be as low as the TeV-scale.  This raises the exciting possibility of probing quantum gravity
effects in high-energy collisions, either at the LHC or in cosmic rays
\cite{Banks:1999gd}.  Of the possible quantum gravity processes, the creation of microscopic black holes would be particularly spectacular
\cite{Dimopoulos:2001hw,Giddings:2001bu,Kanti:2004nr,Landsberg:2003br,Majumdar:2005ba,Park:2012fe,Webber:2005qa}.

If such a microscopic, higher-dimensional, black hole were created,  it will be short-lived, decaying rapidly due to the emission of Hawking radiation.
It is expected that for a significant proportion of the evolution of the black hole
it can be modelled semi-classically \cite{Giddings:2001bu}.
In this regime the geometry of the black hole is regarded as classical, with quantum fields propagating on a background space-time metric.
The details of the Hawking radiation emitted by the black hole as it decays are of particular importance for the simulation of black hole events at the LHC
\cite{Dai:2007ki,Frost:2009cf}.
There is now a vast literature on Hawking radiation from higher-dimensional black holes, see
\cite{Kanti:2004nr,Park:2012fe,Casanova:2005id,Kanti:2008eq,Kanti:2012jh,Winstanley:2007hj,Kanti:2014vsa}
for some reviews on this subject.

The fluxes of energy and angular momentum in the Hawking radiation are just two components of the expectation value of the stress-energy tensor
$\langle {\hat {T}}_{\mu \nu } \rangle $ for a quantum field on a black hole background. The other components of $\langle {\hat {T}}_{\mu \nu }\rangle $ also contain physical information, and the stress-energy tensor as a whole governs the back-reaction of the quantum field on the space-time geometry via the semi-classical Einstein
equations
\begin{equation}
G_{\mu \nu } = 8\pi G \langle {\hat {T}}_{\mu \nu } \rangle .
\label{eq:scee}
\end{equation}
The stress-energy tensor ${\hat {T}}_{\mu \nu }$ involves products of a quantum field operator at the same space-time point and therefore, in general, requires renormalization.
The black holes of phenomenological interest for high-energy collisions either have a single axis of rotation which lies in the brane or are non-rotating.
In both these cases, the symmetries of the brane black hole space-time mean that the components of $\langle {\hat {T}}_{\mu \nu } \rangle $ which comprise the Hawking fluxes of energy and angular momentum on the brane do not require renormalization \cite{Frolov:1989jh,Casals:2005sa,Casals:2008pq}.

Even in four space-time dimensions, computing the full renormalized expectation value of the stress-energy tensor for a quantum field on a background black hole space-time is a complicated process
\cite{Anderson:1993if,Anderson:1994hg,Breen:2011af,Breen:2011aa,Elster:1984hu,Fawcett:1983dk,Groves:2002mh,Howard:1985yg,Howard:1984qp,Jensen:1988rh,Jensen:1992mv,Jensen:1995qv,
Ottewill:2010hr}.
In more than four space-time dimensions, Decanini and Folacci \cite{Decanini:2005eg} have developed a general formalism for the computation of
$\langle {\hat {T}}_{\mu \nu }\rangle _{\text {ren}}$ for a quantum scalar field ${\hat {\phi }} $ based on Hadamard renormalization.
Some general properties of $\langle {\hat {T}}_{\mu \nu }\rangle _{\text {ren}}$ for higher-dimensional black hole space-times are derived in \cite{Morgan:2007hp}, but to date no computation of $\langle {\hat {T}}_{\mu \nu }\rangle _{\text {ren}}$ on the full space-time exterior to a higher-dimensional black hole event horizon has been attempted.
A quantum scalar field is mathematically the simplest type of quantum field, and in this case the renormalized vacuum polarization $\langle {\hat {\phi }} ^{2} \rangle _{\text {ren}}$ is considerably easier to calculate and shares some features with the full renormalized stress-energy tensor.
The vacuum polarization is also of physical significance for spontaneous symmetry breaking (see, for example, \cite{Fawcett:1981fw} for a discussion of this in the black hole context).

The vacuum polarization has been computed on a wide variety of black hole space-times in four dimensions
\cite{Anderson:1989vg,Breen:2010ux,Candelas:1980zt,Candelas:1984pg,Candelas:1985ip,DeBenedictis:1998be,Flachi:2008sr,Frolov:1982pi,Frolov:1982fr,Frolov:1984ra,Ottewill:2010bq,Winstanley:2007tf}.
For higher-dimensional black holes, as well as the approach of \cite{Decanini:2005eg} based on Hadamard renormalization, a general formalism for computing $\langle {\hat {\phi }} ^{2} \rangle _{\text {ren}}$ based on de Witt-Schwinger renormalization has been developed \cite{Thompson:2008bk}.
An exact expression for the renormalized vacuum polarization on the event horizon of a five-dimensional asymptotically flat black hole has been found in \cite{Frolov:1989rv}.
As far as we are aware, the only complete computation of the renormalized vacuum polarization everywhere outside the event horizon of a higher-dimensional black hole is the work of \cite{Shiraishi:1993ti} on a five-dimensional, asymptotically anti-de Sitter black hole.

In this paper we study the vacuum polarization $\langle {\hat {\phi }} ^{2} \rangle _{\text {ren}}$ of a massless, conformally coupled, quantum scalar field on the brane of a higher-dimensional black hole in the context of the AADD brane-world scenario \cite{Antoniadis:1998ig,ArkaniHamed:1998nn,ArkaniHamed:1998rs}.
In this set-up, our universe is a four-dimensional brane in a $D$-dimensional bulk space-time (where $D>4$).
The $D-4$ extra dimensions are flat but compactified.
The radius of compactification is sufficiently small to avoid contradictions with experimental searches for deviations from Newton's Law of Gravitation,
but is typically large compared with the Planck length.
We use a very simple model of a black hole in this scenario, assuming that the brane is tensionless and its thickness negligible.  We also assume that the
size of the black hole is very small compared to the compactification radius of the extra dimensions.
With these assumptions, the black hole can be modelled as a higher-dimensional, asymptotically flat, solution of the vacuum Einstein equations.
For simplicity, we restrict our attention to static, spherically symmetric, higher-dimensional black holes described by the Schwarzschild-Tangherlini metric
\cite{Tangherlini:1963bw}.

In the AADD brane-world scenario, the particles and forces of the Standard Model are constrained to live on the brane, in order to avoid contradictions with
precision particle-physics experiments.  Only gravitational degrees of freedom can propagate in the bulk extra dimensions.
While it is possible to have scalar fields in the gravitational sector of the theory and therefore in the bulk space-time,
studies of Hawking radiation have revealed that the emission of scalar fields in the bulk is suppressed relative to the emission on the brane
\cite{Casals:2008pq,Emparan:2000rs,Harris:2003eg}.
With this in mind, in this paper we consider only the vacuum polarization on the brane.
This means that we can restrict our attention to the four-dimensional on-brane metric describing the black hole, and therefore use established methodology
\cite{Anderson:1994hg,Winstanley:2007tf,Breen:2010ux} to compute $\langle {\hat {\phi }} ^{2} \rangle _{\text {ren}}$.

The outline of this paper is as follows.
In Sec.~\ref{sec:overall} we outline the metric of the black hole on the brane, and construct the point-split Euclidean Green's function which will be used to calculate the vacuum polarization when the quantum scalar field is in the Hartle-Hawking state \cite{Hartle:1976tp}.
We use two different methods for finding $\langle {\hat {\phi }} ^{2} \rangle _{\text {ren}}$ outside and on the event horizon of the brane black hole.
In Sec.~\ref{sec:away} we follow the approach of \cite{Anderson:1994hg,Winstanley:2007tf}, using time-like point-splitting, to find an expression for the vacuum polarization outside the event horizon.
The vacuum polarization on the event horizon is calculated in Sec.~\ref{sec:horizon} using radial point-splitting \cite{Breen:2010ux}.
On the horizon we have a closed-form expression for $\langle {\hat {\phi }} ^{2} \rangle _{\text {ren}}$, but away from the horizon numerical computation is required.
The results of this computation are presented in Sec.~\ref{sec:results} for $D=4\,\ldots ,11$ and our conclusions are in Sec.~\ref{sec:conc}.

\section{Methodology}
\label{sec:method}

\subsection{Set-up}
\label{sec:overall}

We start with the metric for a $D$-dimensional Schwarzschild-Tangherlini \cite{Tangherlini:1963bw} black hole:
\begin{equation}
ds^{2} = -f(r) \, dt^{2} + f(r) ^{-1} dr^{2}+ r^{2} \, d\Omega _{D-2}
\label{eq:bulk}
\end{equation}
where
\begin{equation}
f(r) = 1- \left( \frac {r_{h}}{r} \right) ^{D-3}
\label{eq:metricf}
\end{equation}
and $d\Omega _{D-2}$ is the line element on the $(D-2)$-sphere $S^{D-2}$.
Here and throughout this paper we use Lorentzian metric signature $(-,+,\ldots, + )$ and units in which
$8\pi G = c = \hbar = k_{B}=1$.
The metric (\ref{eq:bulk}) represents a static, spherically symmetric,  black hole with mass
\begin{equation}
M = \frac {1}{2}\left( D-2 \right) r_{h}^{D-3} A_{D-2}
\end{equation}
where $A_{D-2}=2\pi ^{(D-1)/2}/\Gamma [ (D-1)/2]$, with $\Gamma $ the Euler gamma function, is the area of the $(D-2)$-sphere.

Labelling the angular co-ordinates on the $(D-2)$-sphere above by $\theta$, $\varphi $, $\theta _{3},\ldots ,\theta _{D-2}$, the metric of the brane black hole is constructed by setting $\theta _{i}=\pi /2$ for $i=3,\ldots , D-2$, giving the on-brane metric
\begin{equation}
ds^{2} = - f(r) \, dt^{2} + f(r)^{-1} dr^{2} + r^{2} \left[ d\theta ^{2} + \sin ^{2} \theta \, d\varphi ^{2} \right] .
\label{eq:brane}
\end{equation}
In the four-dimensional brane metric (\ref{eq:brane}), the function $f(r)$ still has the form (\ref{eq:metricf}), and so depends on the number of bulk space-time dimensions $D$.
The brane metric (\ref{eq:brane}) is therefore not a solution of the vacuum Einstein equations except in the case $D=4$, when it reduces to the usual Schwarzschild form.
When $D>4$, the metric (\ref{eq:brane}) is a solution of the nonvacuum Einstein equations with an effective fluid source on the brane \cite{Sampaio:2009ra}.

In this paper we consider a massless quantum scalar field ${\hat {\phi }} $ with conformal coupling to the four-dimensional space-time geometry (\ref{eq:brane}). The scalar field satisfies the Klein-Gordon equation
\begin{equation}
\left[ \nabla _{\mu } \nabla ^{\mu } - \xi R \right] {\hat {\phi }} =0,
\end{equation}
where $\xi = 1/6$ is the coupling constant for conformal coupling on the four-dimensional brane black hole space-time (\ref{eq:brane}) and $R$ is the Ricci scalar of the four-dimensional on-brane metric (\ref{eq:brane}), which takes the form
\begin{equation}
R = \frac {(D-4)(D-5)r_{h}^{D-3}}{r^{D-1}}.
\label{eq:Ricci}
\end{equation}
The Ricci scalar therefore vanishes for $D=4, 5$.

To compute the renormalized vacuum polarization, we follow \cite{Anderson:1993if,Anderson:1994hg,Winstanley:2007tf} and take a Euclidean approach.
Defining Euclidean time $\tau = it$, the brane metric (\ref{eq:brane}) becomes
\begin{equation}
ds^{2} = f(r) \, d\tau ^{2}  + f(r)^{-1} dr^{2} + r^{2} \left[ d\theta ^{2} + \sin ^{2} \theta \, d\varphi ^{2} \right] .
\label{eq:Emetric}
\end{equation}
The point-split Euclidean Green's function $G_{E}(x;x')$ satisfies the equation \cite{Anderson:1993if,Anderson:1994hg}
\begin{equation}
\left[ \nabla _{x\mu }\nabla _{x}^{\mu } - \xi R \right] G_{E}(x;x') = -g^{-1/2} (x) \delta ^{4} (x,x')
\label{eq:KG}
\end{equation}
where the covariant derivative is taken with respect to the Euclidean metric (\ref{eq:Emetric}), which has determinant $g$.
The unrenormalized expectation value of the vacuum polarization $\langle {\hat {\phi }} ^{2} \rangle _{\text {unren}}$ is given by the coincidence limit
\begin{equation}
\langle {\hat {\phi }} ^{2} \rangle _{\text {unren}} = \Re \left[ \lim _{x\rightarrow x'} G_{E}(x;x') \right] .
\label{eq:unren}
\end{equation}

We consider the quantum scalar field to be in the thermal Hartle-Hawking state \cite{Hartle:1976tp} at temperature $T$.  The four-dimensional black hole (\ref{eq:brane}) has Hawking temperature
\begin{equation}
T = \frac {D-3}{4\pi r_{h}}
\label{eq:temp}
\end{equation}
which increases linearly with $D$ for fixed $r_{h}$, and is inversely proportional to $r_{h}$ for fixed $D$.
For a thermal state, the point-split scalar Euclidean Green's function $G_{E}(x;x')$ is periodic in $\tau - \tau '$ with period $T^{-1}$, and takes the form
\cite{Anderson:1989vg,Anderson:1993if,Anderson:1994hg}
\begin{eqnarray}
G_{E}(x;x') & = &
\frac {T}{4\pi } \sum _{n=-\infty }^{\infty } \exp \left[ i \omega \left( \tau - \tau' \right)  \right]
\nonumber \\
 & & \times \sum _{\ell = 0}^{\infty } \left( 2\ell + 1 \right) P_{\ell }\left( \cos \gamma \right) G_{\omega \ell }(r;r') \qquad
 \label{eq:GEgen}
\end{eqnarray}
where $\omega = 2n\pi T$, and $P_{\ell }$ is the usual Legendre function with
\begin{equation}
\cos \gamma = \cos \theta \cos \theta' + \sin \theta \sin \theta' \cos \left( \varphi - \varphi' \right) .
\label{eq:gamma}
\end{equation}
The radial Green's function $G_{\omega \ell }$ takes the form
\begin{equation}
G_{\omega \ell }(r,r') = C_{\omega \ell } p_{\omega \ell }(r_{<}) q_{\omega \ell }(r_{>})
\label{eq:Y}
\end{equation}
where $p_{\omega \ell }(r)$ and $q_{\omega \ell }(r)$ are solutions of the radial equation
\begin{eqnarray}
0 & = &
f \frac {d^{2}G_{\omega \ell }}{dr^{2}} + \left( \frac {2f}{r} + \frac {df}{dr} \right) \frac {dG_{\omega \ell }}{dr}
\nonumber \\ &  &
-\left[ \frac {\omega ^{2}}{f} + \frac {\ell \left( \ell + 1 \right)}{r^{2}} + \frac {R}{6} \right] G_{\omega \ell }
\label{eq:radial}
\end{eqnarray}
which is the homogeneous version of the radial equation for $G_{\omega \ell }$ arising from separating the inhomogeneous Klein-Gordon equation
(\ref{eq:KG}) on the metric (\ref{eq:Emetric}).
The function $p_{\omega \ell }(r)$ is defined as the solution of (\ref{eq:radial}) which is regular at the event horizon $r=r_{h}$,
while $q_{\omega \ell }(r)$
is the solution of (\ref{eq:radial}) which is regular as $r\rightarrow \infty $.
In (\ref{eq:Y}), as usual, $r_{<}$ is the smaller of the two values $r$, $r'$, while $r_{>}$ is the greater.
The normalization constant $C_{\omega \ell }$ in (\ref{eq:Y}) is determined by the normalization condition \cite{Anderson:1993if,Anderson:1994hg}
\begin{equation}
C_{\omega \ell } \left[ p_{\omega \ell } \frac {dq_{\omega \ell }}{dr}
- q_{\omega \ell }\frac {dp_{\omega \ell }}{dr} \right]
= -\frac {1}{r^{2}f}.
\label{eq:C}
\end{equation}

Before we can bring the space-time points together in (\ref{eq:unren}) by taking the limit $x\rightarrow x'$, we need to subtract divergent terms.  For a massless, conformally coupled scalar field, these take the simple form \cite{Christensen:1978yd}
\begin{equation}
\langle {\hat {\phi }} ^{2} \rangle _{\text {div}} = \frac {1}{8\pi ^{2}\sigma } + \frac {1}{96\pi ^{2}}
\frac {R_{\alpha \beta } \sigma ^{\alpha }\sigma ^{\beta }}{\sigma } ,
\label{eq:div}
\end{equation}
where $\sigma $ is one-half the square of the geodesic distance between the points $x$ and $x'$ and $\sigma ^{\alpha }=\sigma ^{;\alpha }$
with $R_{\alpha \beta }$ the Ricci tensor.
The detailed form of the divergent terms (\ref{eq:div}) depends on the point-splitting chosen.

For the remainder of this section, our methodology for computing the renormalized vacuum polarization falls into two parts: (a) the computation away from the event horizon, using temporal point-splitting following
\cite{Anderson:1994hg}, and (b) the computation on the event horizon, using radial point-splitting following \cite{Breen:2010ux}.
In the rest of this paper we set the event horizon radius $r_{h}=1$, so that all lengths having numerical values are
in units of the event horizon radius.

\subsection{Outside the event horizon}
\label{sec:away}

Outside the event horizon, we follow the standard methodology developed in \cite{Anderson:1994hg} to compute the renormalized vacuum polarization
$\langle {\hat {\phi }}^{2} \rangle _{\text {ren}}$.
We choose temporal point-splitting, setting $r=r'$, $\theta = \theta '$ and $\varphi =\varphi '$.
Then $\cos \gamma =1 $ (\ref{eq:gamma}) and using the fact that $P_{\ell }(1)=1$, the Euclidean Green's function (\ref{eq:GEgen}) takes the form
\begin{eqnarray}
G_{E}(\tau, {\bmath {x}}; \tau' , {\bmath {x}}') & =&
\nonumber \\ & &
\hspace{-3cm} \frac {T}{4\pi } \sum _{n=-\infty }^{\infty } e^{i\omega \epsilon } \sum _{\ell =0}^{\infty } \left( 2\ell + 1\right)
 C_{\omega \ell }p_{\omega \ell }(r) q_{\omega \ell }(r),
\label{eq:tsplit}
\end{eqnarray}
where $\epsilon = \tau - \tau'$.
As is well known, the sums over $\ell $ in (\ref{eq:tsplit}) do not converge although, by definition, the Euclidean Green's function $G_{E}$ must be finite
when the points are separated.
This is remedied by subtracting from (\ref{eq:tsplit}) a suitable multiple of the Dirac delta function, which vanishes when the points do not coincide,
giving \cite{Anderson:1994hg}
\begin{eqnarray}
G_{E} (\tau, {\bmath {x}}; \tau' , {\bmath {x}}') & =&
\nonumber \\  & & \hspace{-3cm}
\frac {T}{4\pi } \sum _{n=-\infty }^{\infty } e^{i\omega \epsilon } \sum _{\ell =0}^{\infty } \left[ \left( 2\ell + 1\right)
C_{\omega \ell }p_{\omega \ell }(r) q_{\omega \ell }(r) - \frac {1}{r{\sqrt {f}}} \right] .
\nonumber \\
\label{eq:tsplitfinal}
\end{eqnarray}
For temporal point-splitting, the divergent terms (\ref{eq:div}) take the following form \cite{Anderson:1994hg} for a massless, conformally coupled, scalar field:
\begin{eqnarray}
\langle {\hat {\phi }} ^{2} \rangle _{\text {div}} & = &
\frac {1}{4\pi ^{2}f\epsilon ^{2}}
+ \frac {1}{192\pi ^{2} f} \left( \frac {df}{dr} \right) ^{2} - \frac {1}{96\pi ^{2}} \frac {d^{2}f}{dr^{2}}
\nonumber \\ & &
- \frac {1}{48\pi ^{2}r}\frac {df}{dr}.
\end{eqnarray}
These divergent terms can be written as mode sums by using standard identities \cite{Anderson:1994hg}, subtracted from (\ref{eq:tsplitfinal}) and then the limit $\epsilon \rightarrow 0$ taken.
This gives the final renormalized expectation value to be \cite{Anderson:1994hg}
\begin{equation}
\langle {\hat {\phi }} ^{2} \rangle _{\text {ren}} = \langle {\hat {\phi }} ^{2} \rangle _{\text {analytic}}
+ \langle {\hat {\phi }} ^{2} \rangle _{\text {numeric}}
\end{equation}
where
\begin{subequations}
\begin{eqnarray}
\langle {\hat {\phi }} ^{2}\rangle _{\text {analytic}}
& = &
\frac {T^{2}}{12f} - \frac {1}{192\pi ^{2} f} \left( \frac {df}{dr} \right) ^{2} + \frac {1}{96\pi ^{2}} \frac {d^{2}f}{dr^{2}}
\nonumber \\ & &
+ \frac {1}{48\pi ^{2}r}\frac {df}{dr},
\label{eq:analytic}
\\
 \langle {\hat {\phi }} ^{2} \rangle _{\text {numeric}}
& = &  \nonumber \\ & &
\hspace{-2cm} \frac {T}{2\pi } \sum _{n=1}^{\infty } \left\{ \sum _{\ell =0}^{\infty } \left[ \left( 2\ell + 1\right) C_{\omega \ell }p_{\omega \ell }(r)
q_{\omega \ell }(r)
- \frac {1}{r{\sqrt {f}}} \right] + \frac {\omega }{f} \right\}
\nonumber \\ & &
\hspace{-1.5cm}
+\frac {T}{4\pi } \sum _{\ell = 0}^{\infty } \left[ \left( 2\ell + 1 \right) C_{0\ell }p_{0\ell }(r) q_{0\ell }(r) - \frac {1}{r{\sqrt {f}}} \right] .
\nonumber \\ & &
\label{eq:numeric}
\end{eqnarray}
\end{subequations}

The mode functions $p_{\omega \ell }(r)$, $q_{\omega \ell }(r)$ cannot be found in closed form for $\omega >0$, so that, as the name suggests,
$\langle {\hat {\phi }} ^{2} \rangle _{\text {numeric}}$ (\ref{eq:numeric}) is computed numerically while we have a closed form expression for
$\langle {\hat {\phi }} ^{2}\rangle _{\text {analytic}}$ (\ref{eq:analytic}).

To speed up the convergence of the mode sums in (\ref{eq:numeric}), we follow
\cite{Winstanley:2007tf} and subtract a WKB-like approximation, which encodes the large $\omega$, $\ell $ behaviour of the mode functions.
We define a new function $\zeta _{\omega \ell }(r)$ {\footnote{In the literature this quantity is usually denoted $\beta _{\omega \ell}$ \cite{Winstanley:2007tf,Casals:2009xa} but here we use the notation $\zeta _{\omega \ell }$ to avoid confusion with $\beta _{\ell }$ introduced in Sec.~\ref{sec:horizon}.}}  by
\begin{equation}
\zeta _{\omega \ell }(r)= C_{\omega \ell }p_{\omega \ell }(r)q_{\omega \ell }(r),
\end{equation}
which satisfies the following differential equation \cite{Howard:1985yg}
\begin{eqnarray}
\frac {1}{\zeta _{\omega \ell} ^{2}} & = &
\nonumber \\ & & \hspace{-1.8cm}
4\chi _{\omega \ell }^{2} \left\{ 1 - \frac {1}{\chi _{\omega \ell }^{2}}
\left[ \frac {r^{2}f}{{\sqrt {\zeta _{\omega \ell }}}} \frac {d}{dr}  \left( r^{2}f \frac {d\left( {\sqrt {\zeta _{\omega \ell }}}\right) }{dr} \right)
- \eta \right] \right\},
\label{eq:WKB1}
\end{eqnarray}
where
\begin{subequations}
\begin{eqnarray}
\chi _{\omega \ell }(r) & = &
{\sqrt {\omega ^{2}r^{4} + \left( \ell + \frac {1}{2} \right) ^{2} r^{2} f }},
\\
\eta (r) & = & \frac {R}{6}fr^{4} -\frac {1}{4} fr^{2} ,
\end{eqnarray}
and the Ricci scalar $R$ is given by (\ref{eq:Ricci}).
\end{subequations}
The WKB expansion is found by inserting a fictitious parameter $\varepsilon $ in (\ref{eq:WKB1}):
\begin{eqnarray}
\frac {1}{\zeta _{\omega \ell} ^{2}} & = &
\nonumber \\  & & \hspace{-1.5cm}
4\chi _{\omega \ell }^{2} \left\{ 1 - \frac {1}{\varepsilon ^{2} \chi _{\omega \ell }^{2}}
\left[ \frac {r^{2}f}{{\sqrt {\zeta _{\omega \ell }}}} \frac {d}{dr}  \left( r^{2}f \frac {d\left( {\sqrt {\zeta _{\omega \ell }}}\right) }{dr} \right)
- \eta \right] \right\},
\nonumber \\ & &
\label{eq:WKB2}
\end{eqnarray}
expanding $\zeta _{\omega \ell }(r)$ in inverse powers of $\varepsilon $
\begin{equation}
\zeta _{\omega \ell } = \zeta _{0\omega \ell }(r) + \varepsilon ^{-2} \zeta _{1\omega \ell }(r) + \varepsilon ^{-4} \zeta _{2\omega \ell }(r)
+ \varepsilon ^{-6} \zeta _{3\omega \ell }(r) + \ldots ,
\label{eq:expansion}
\end{equation}
and then setting $\varepsilon =1$ at the end of the calculation.
The WKB terms $\zeta _{i\omega \ell }(r)$ have the form
\begin{equation}
\zeta _{i\omega \ell }(r) = \sum _{k=1}^{2i+1} A_{i,k}(\omega, r) \chi _{\omega \ell } ^{-\left( 2i + 2k -1\right) }
\label{eq:WKBterms}
\end{equation}
where the functions $A_{i,k}(\omega ,r)$ depend on $\omega $ (and $r$) but not $\ell $.
The first of these functions is $A_{0,1}(\omega ,r)=1/2$ and the forms of $A_{1,k}(\omega ,r)$ can be found in \cite{Winstanley:2007tf}.
To reduce the number of mode functions $p_{\omega \ell }(r)$, $q_{\omega \ell }(r)$ which have to be found numerically, we used an expansion
(\ref{eq:expansion}) up to and including $\zeta _{3\omega \ell }(r)$; the remaining functions $A_{2,k}(\omega ,r)$ and $A_{3,k}(\omega ,r)$ can be found in \cite{hewitt}.

Adding and subtracting the WKB expansion from (\ref{eq:numeric}), we arrive at the expression
\begin{widetext}
\begin{eqnarray}
\langle {\hat {\phi }} ^{2} \rangle _{\text {numeric}} & = &
\frac {T}{2\pi } \sum _{n=1}^{\infty } \left\{ \sum _{\ell =0}^{\infty } \left( 2 \ell + 1 \right) \left[
C_{\omega \ell } p_{\omega \ell }(r) q_{\omega \ell }(r) - \zeta _{0\omega \ell }(r) - \zeta _{1\omega \ell }(r)
-\zeta _{2\omega \ell }(r) - \zeta _{3\omega \ell }(r) \right]
\right. \nonumber \\ & & \left.
+ \sum _{\ell =0}^{\infty } \left( 2 \ell + 1 \right) \left[ \zeta _{0\omega \ell }(r) + \zeta _{1\omega \ell }(r)
+ \zeta _{2\omega \ell }(r) +\zeta _{3\omega \ell }(r) - \frac {1}{r{\sqrt {f}}} \right]
+ \frac {\omega }{f}\right\}
\nonumber \\ & &
+ \frac {T}{4\pi } \sum _{\ell = 0}^{\infty }
\left( 2\ell + 1 \right) \left[ C_{0\ell } p_{0\ell }(r) q_{0\ell }(r) - \zeta _{00\ell }(r) - \zeta _{10\ell }(r) - \zeta _{20\ell }(r)
-\zeta _{30\ell }(r) \right]
+ \Delta _{1} + \Delta _{2} + \Delta _{3},
\label{eq:phi2numnext}
\end{eqnarray}
\end{widetext}
where
\begin{eqnarray}
\Delta _{1} & = & \sum _{\ell =0}^{\infty } \left( 2 \ell + 1 \right) \zeta _{10\ell }(r) = \frac {\pi ^{2}}{r^{3} f^{\frac {3}{2}}} A_{1,1}(0,r),
\nonumber \\
\Delta _{2} & = & \sum _{\ell =0}^{\infty } \left( 2 \ell + 1 \right) \zeta _{20\ell  }(r) = \frac {\pi ^{4}}{3r^{5}f^{\frac {5}{2}}} A_{2,1}(0,r),
\nonumber \\
\Delta _{3} & = & \sum _{\ell =0}^{\infty } \left( 2 \ell + 1 \right) \zeta _{30\ell }(r) = \frac {2\pi ^{6}}{15r^{7}f^{\frac {7}{2}}} A_{3,1}(0,r).
\qquad
\label{eq:Delta}
\end{eqnarray}
The mode sums in the first and third lines of (\ref{eq:phi2numnext}) are now rapidly converging.  It remains to find the sums over the WKB expansions in the second line of (\ref{eq:phi2numnext}).
To do this, we follow \cite{Candelas:1984pg,Winstanley:2007tf,Jensen:1988rh} and employ the Watson-Somerfeld identity
\begin{eqnarray}
\sum _{\ell =0}^{\infty } {\mathcal {F}}(\ell ) & = &
\int _{\lambda =0}^{\infty } {\mathcal {F}} \left( \lambda - \frac {1}{2} \right) \, d\lambda
\nonumber \\ & &
- \Re \left[ i \int _{\lambda =0}^{\infty } \frac {2}{1+e^{2\pi \lambda }} {\mathcal {F}} \left( i\lambda - \frac {1}{2} \right) \, d\lambda \right]
\nonumber \\ & &
\label{eq:WS}
\end{eqnarray}
which is valid for any function ${\mathcal {F}}(\ell )$ analytic in the right-hand half-plane.
Using (\ref{eq:WS}) we write \cite{Winstanley:2007tf} (for $\omega >0$)
\begin{eqnarray}
\sum _{\ell =0}^{\infty } \left( 2\ell + 1 \right) \left[ \zeta _{0\omega \ell } (r) - \frac {1}{r{\sqrt {f}}} \right]
& = & I_{0} (\omega ,r) + J_{0}(\omega ,r)
\nonumber \\ & &
 + \frac {1}{24\omega r^{2}} ,
\nonumber \\
\sum _{\ell =0}^{\infty } \left( 2\ell + 1 \right) \zeta _{k\omega \ell }(r)
& = & I_{k}(\omega ,r) + J_{k}(\omega ,r) ,
\nonumber \\ & &
\end{eqnarray}
for $k=1,2,3$, where the $I_{k}(\omega ,r)$ are the first integrals coming from (\ref{eq:WS}) and the $J_{k}(\omega ,r)$ are the second.

The $I_{k}(\omega ,r)$ integrals are readily computed; the first two take the following form for a massless, conformally coupled scalar field \cite{Winstanley:2007tf}
\begin{eqnarray}
I_{0}(\omega, r) & = &
\int _{\lambda = 0}^{\infty } \left[ 2\lambda \zeta _{0\omega \ell } - \frac {1}{r{\sqrt {f}}} \right] d\lambda = - \frac {\omega }{f},
\nonumber \\
I_{1}(\omega ,r) & = &
\int _{\lambda =0}^{\infty } 2\lambda \zeta _{1\omega \ell } \, d\lambda =
-\frac {1}{24\omega r^{2}}.
\end{eqnarray}
The remaining $I_{2}(\omega ,r)$ and $I_{3}(\omega ,r)$ are straightforward to find, for example, in {\texttt{Mathematica}}
but the resulting expressions are sufficiently long that we do not include them here \cite{hewitt}.

To compute the $J_{k}$ integrals, we first make a change of variables \cite{Winstanley:2007tf}:
\begin{equation}
\lambda = \rho q, \qquad \rho =\frac {\omega r}{{\sqrt {f}}},
\end{equation}
The integral $J_{0}$ is analysed in \cite{Winstanley:2007tf}. After an integration by parts, it takes the form \cite{Winstanley:2007tf}
\begin{eqnarray}
J_{0} & = &
\frac {\omega }{f} - \frac {1}{24\omega r^{2}} - \frac {4\pi \omega }{f} \int _{\lambda =0}^{\rho }
\left( 1 - \frac {\lambda ^{2}}{\rho ^{2}} \right) ^{\frac {1}{2}} \frac {e^{2\pi \lambda }}{\left( 1 + e^{2\pi \lambda } \right) ^{2}} d\lambda
\nonumber \\ & &
\end{eqnarray}
which is $O(\omega ^{-3})$ as $\omega \rightarrow \infty $.
For the remaining $J_{k}$ integrals, using (\ref{eq:WKBterms}), we have
\begin{eqnarray}
J_{k} & = &
\sum _{j=1}^{2k+1} 4\rho ^{2} \left( \rho ^{2}r^{2}f\right) ^{-y}  A_{k,j} (\omega ,r)
\nonumber \\ & &
\qquad \times \Re \left[
\int _{q=0}^{\infty } \frac {q\left( 1- q^{2} \right) ^{-y} }{1+e^{2\pi \rho q}} dq
\right] ,
\label{eq:Jint1}
\end{eqnarray}
for $k=1,2,3$ where
\begin{equation}
y = \left( 2j +2k -1 \right) /2.
\end{equation}
In \cite{Winstanley:2007tf}, the integral over $q$ in (\ref{eq:Jint1}) was performed by repeated integrations by parts.
Here we take an alternative approach, first defining
\begin{equation}
g(q) = \frac {q}{1+e^{2\pi \rho q}}
\end{equation}
and letting $g_{y}(q)$ be the Taylor series expansion of $g(q)$ about $q=1$ up to order $y+(1/2)$.
Then, for $k=1,2,3$,
\begin{widetext}
\begin{equation}
J_{k} =
\sum _{j=1}^{2k+1} 4\rho ^{2} \left( \rho ^{2}r^{2}f\right) ^{-y}  A_{k,j} (\omega ,r) \, \left\{
\int _{q=0}^{1} \left[ g(q) - g_{y}(q) \right] \left( 1 - q^{2} \right) ^{-y} dq
+
\Re \left[
\int _{q=0}^{\infty } g_{y}(q) \left( 1- q^{2} \right) ^{-y} dq
\right] \right\} .
\label{eq:Jint2}
\end{equation}
The integrand in the first integral in (\ref{eq:Jint2}) is regular at $q=1$ and therefore the first integral is straightforward to compute numerically.
The integrand in the second integral in (\ref{eq:Jint2}) has a pole at $q=1$. Integrating around the pole using the contour shown in Fig.~1 in \cite{Winstanley:2007tf}
is routine in, for example, {\texttt{Mathematica}} as $g_{y}(q)$ is a polynomial in $q$, and yields a finite answer in each case.

Finally, $\langle {\hat {\phi }} ^{2} \rangle _{\text{numeric}}$ defined by Eq.~(\ref{eq:phi2numnext}) takes the form
\begin{eqnarray}
\langle {\hat {\phi }} ^{2} \rangle _{\text {numeric}} & = &
\frac {T}{2\pi } \sum _{n=1}^{\infty } \bigg\{ \sum _{\ell =0}^{\infty } \left( 2 \ell + 1 \right) \left[
C_{\omega \ell } p_{\omega \ell }(r) q_{\omega \ell }(r) - \zeta _{0\omega \ell }(r) - \zeta _{1\omega \ell }(r)
-\zeta _{2\omega \ell }(r) - \zeta _{3\omega \ell }(r) \right]
 \nonumber \\ & &
+ I_{2}+ I_{3}
+ J_{0} + J_{1} + J_{2} + J_{3}
\bigg\}
\nonumber \\ & &
+ \frac {T}{4\pi } \sum _{\ell = 0}^{\infty }
\left( 2\ell + 1 \right) \left[ C_{0\ell } p_{0\ell }(r) q_{0\ell }(r) - \zeta _{00\ell }(r) - \zeta _{10\ell }(r) - \zeta _{20\ell }(r)
-\zeta _{30\ell }(r) \right]
+ \Delta _{1} + \Delta _{2} + \Delta _{3}.
\label{eq:numericfinal}
\end{eqnarray}
The results of numerically computing (\ref{eq:numericfinal}) will be discussed in Sec.~\ref{sec:numeric}.
\end{widetext}

\subsection{On the horizon}
\label{sec:horizon}

The method discussed in the previous subsection works well for any $r>1$ but is not well-suited to a computation of $\langle {\hat {\phi }} ^{2} \rangle _{\text {ren}}$
on the horizon $r=1$ (for example, there are terms in the $\Delta _{i}$ (\ref{eq:Delta}) which diverge as $r\rightarrow 1$ and $f\rightarrow 0$).
We therefore take a different approach to computing $\langle {\hat {\phi }} ^{2}\rangle _{\text {ren}}$ on the horizon, following \cite{Breen:2010ux} and
using radial instead of temporal point-splitting.

Separating the space-time points along the radial direction, we may express the unrenormalized vacuum polarization (\ref{eq:unren}) in the form
\begin{eqnarray}
\langle \hat{\phi }^2\rangle_{\text{unren}} & = &
\nonumber \\ & & \hspace{-2.5cm}
\lim_{r \to r'}\left[ \sum_{n=0}^{\infty} F(n)\sum_{\ell=0}^{\infty}\left( 2\ell+1 \right)
C_{\omega \ell }p_{\omega \ell }(r_{<})q_{\omega \ell }(r_{>}) \right] ,
\label{eq:unrenhor}
\end{eqnarray}
where  $r_<=\min (r,r')$, $r_>=\max (r,r')$, and
\begin{equation}
F(n) =
\begin{cases}
T/4\pi,  & \qquad n=0, \\
T/2\pi,  & \qquad n>0 .
\end{cases}
\end{equation}
We now set $r_>=r$, place the inner point, $r'$, on the horizon and consider the limit $r\to 1$.
In a neighbourhood of $r'=1$, the solution $p_{\omega \ell }(r')$ of the radial equation (\ref{eq:radial}) which is regular at the horizon possesses the series expansion \cite{Breen:2010ux}
\begin{equation}
p_{\omega \ell }(r')=\frac{1}{\sqrt{2\pi T}} \left( r'-1 \right) ^{{\frac {n}{2}}} +  O\left[ \left( r'-1 \right) ^{1+\frac {n}{2}} \right] ,
\end{equation}
where $\omega = 2\pi n T$.
Setting $r'=1$ therefore has the effect that $p_{\omega \ell }(r')$ vanishes for $n>0$. Therefore (\ref{eq:unrenhor})  reduces to:
\begin{equation}
\langle \hat{\phi }^2\rangle_{\text{unren}}=\lim_{r \to 1}{\sqrt {\frac {T}{32\pi ^{3}}}}\sum_{\ell =0}^{\infty}
\left( 2\ell +1 \right) C_{0\ell }q_{0\ell }(r).
\label{eq:unren2}
\end{equation}
The advantage of using radial point separation is now clear as we have a sum which involves only the zero frequency modes. In fact, as argued in \cite{Breen:2011aa}, we may only choose radial or angular separation for on-horizon calculations as temporal separation is meaningless there.

We renormalize (\ref{eq:unren2}) using the method of  Brown and Ottewill \cite{Brown:1986tj}, by subtracting off the singular terms in the Hadamard expansion of the Euclidean Green's function, which we denote by  $\langle {\hat {\phi }}^{2} \rangle _{\text {div}}$. These have been worked out explicitly for radial point separation in \cite{Breen:2010ux}, and for a conformally coupled massless scalar field (with $f(r)$ given by (\ref{eq:metricf}) with $r_{h}=1$) take the form
\begin{equation}
\langle {\hat {\phi }}^{2} \rangle _{\text {div}}=
\frac{D-3}{16 \pi^2 (r-1)} -\frac{D-3}{48\pi^2 }\\
+O\left[ (r-1) \ln(r-1) \right].
\end{equation}
We then have a formal expression for $\langle \hat{\phi }^2\rangle_{\text{ren}}$, given by
\begin{eqnarray}
\langle \hat{\phi }^2\rangle_{\text {ren}} & = &
\lim_{r \to 1} \left[ {\sqrt {\frac {T}{32\pi ^{3}}}} \sum_{\ell=0}^{\infty}\left( 2\ell+1 \right) C_{0\ell}q_{0\ell}(r)
\right.
\nonumber\\
& &  \left.  -\frac{D-3}{16 \pi^2 (r-1)} +\frac{D-3}{48\pi^2 }\right] .
\label{eq:horizonfinal}
\end{eqnarray}

We now need to calculate the sum over the modes $q_{0\ell }(r)$.
We begin this process by noting that the radial equation (\ref{eq:radial}), for the $n=0$ modes, possesses closed-form solutions in terms of hypergeometric functions. To demonstrate this, we define new independent and dependent variables as follows:
\begin{equation}
x=1-r^{D-3}, \qquad
Z(x)= (1-x)^{\alpha } G_{0\ell }
\end{equation}
with
\begin{equation}
\alpha=\frac{3(4-D) -\sqrt{3(D-4)(D-2)}}{6(D-3)} .
\label{eq:alpha}
\end{equation}
Under this transformation,  (\ref{eq:radial}) takes the form
\begin{eqnarray}
0 & = & x(1-x)Z''(x)
\nonumber \\ & &
+  \left[
1-\left(2+\frac{\sqrt{(D-4)(D-2)}}{\sqrt{3}(D-3)}\right)x\right] Z'(x)
\nonumber \\ & &
+ \left(\frac{\ell}{D-3} +\alpha\right)\left(\frac{\ell+1}{D-3} -\alpha\right)  Z(x).
\label{eq:modetrans}
\end{eqnarray}
We may express (\ref{eq:modetrans}) in the form of the hypergeometric differential equation \cite{DLMF},
\begin{equation}
x(1-x)Z''(x) +\left[ {\mathfrak {c}}-({\mathfrak {a}}+{\mathfrak {b}}+1)x\right] Z'(x) -{\mathfrak {ab}} Z(x) =0,
\end{equation}
by making the following identifications
\begin{equation}
{\mathfrak {a}}=-\frac{\ell}{D-3} -\alpha, \qquad  {\mathfrak {b}}=\frac{\ell+1}{D-3} -\alpha, \qquad {\mathfrak {c}}=1.
\end{equation}

\begin{widetext}
It then follows that (\ref{eq:modetrans}) has the following pair of linearly independent
solutions
\begin{eqnarray}
Z_{1}(x) & = &
{}_2 F_1\left( -\frac{\ell}{D-3} -\alpha , \frac{\ell+1}{D-3} -\alpha ;
1; x \right) ,
\nonumber \\
Z_{2}(x) & = &
(1-x)^{-\frac{\ell+1}{D-3} +\alpha}
{}_2 F_1\left( \frac{\ell+D-3}{D-3} +\alpha, \frac{\ell+1}{D-3} -\alpha ;
\frac{2\ell+D-2}{D-3};\frac{1}{1-x}\right).
\end{eqnarray}
Transforming back to our original variables, and identifying the solutions regular at $r=1$ and $r=\infty$, we find that the two independent solutions to  (\ref{eq:radial}), $p_{0\ell }(r)$ and $q_{0\ell }(r)$, are given by:
\begin{eqnarray}
p_{0\ell}(r) & = &
\frac {1}{\sqrt {2\pi T}} r^{-(D-3)\alpha }{}_2 F_1\left(-\frac{\ell}{D-3} -\alpha , \frac{\ell+1}{D-3} -\alpha ; 1 ; 1-r^{D-3}\right) ,
\nonumber \\
q_{0\ell}(r) & = &
r^{-(\ell+1)}\frac{\Gamma\left(\frac{\ell+D-3}{D-3} +\alpha\right)\Gamma\left({\frac{\ell+1}{D-3} -\alpha}\right)}{2 \sqrt{2\pi T}\Gamma\left(\frac{2\ell+D-2}{D-3}\right)}
{}_2 F_1\left(\frac{\ell+D-3}{D-3} +\alpha , \frac{\ell+1}{D-3} -\alpha ; \frac{2\ell+D-2}{D-3} ; r^{3-D}\right) .
\label{eq:qsol}
\end{eqnarray}
\end{widetext}
The numerical factors in front of $p_{0\ell }(r)$ and $q_{0\ell }(r)$ are chosen so that their series solutions are in agreement with the expressions in \cite{Breen:2010ux}, but this is purely a matter of convention. We could alternatively have  chosen to include these factors into the constant $C_{0\ell}$
(which in our convention is equal to unity).

Returning now to (\ref{eq:horizonfinal}), while we  have a closed form expression for $q_{0\ell}(r)$ given above (\ref{eq:qsol}), unfortunately we are unable to immediately perform the mode sum as, to the best of our knowledge, no closed-form expression for this sum is known.
We therefore adapt the method developed in \cite{Breen:2010ux} to the case at hand.
The procedure in \cite{Breen:2010ux} can be applied to calculate $\langle \hat{\phi }^2\rangle_{\text{ren}}$  for a general spherically symmetric black hole space-time (\ref{eq:Emetric}), where the only constraint on the function $f(r)$ is that it has a single zero at the horizon in question, as is the case here.

We begin by writing the full solution for $q_{0\ell }(r)$ in terms of an approximation in the following manner
\begin{equation}
q_{0\ell}(r)= Q_{0\ell}(r) +\beta_{\ell} p_{0\ell}(r) + \mathcal{R}_{\ell}(r) ,
\label{eq:qapprox}
\end{equation}
where $Q_{0\ell}(r)$ is some approximation to $q_{0\ell}(r)$,  $\beta_{\ell} $ is a function of $\ell$ only and $\mathcal{R}_{\ell}(r)$ denotes the remainder of $q_{0\ell}(r)$ not captured by the first two terms.
The method of \cite{Breen:2010ux} makes use of a uniform approximation, developed using extended Green-Liouville asymptotic analysis.
This gives $Q_{0\ell }(r)$ to be
\begin{equation}
Q_{0\ell}(r)=\left( \frac{\xi(r)}{r^2 f(r)}\right)^{1/4}K_0\left[k_0 \xi^{1/2}(r)\right]
\label{eq:EGL}
\end{equation}
where
\begin{equation}
k_{0}=\sqrt{(\ell+a)(\ell+\bar{a})}
\label{eq:k0}
\end{equation}
with
\begin{equation}
a=\frac{1}{2}\left(1 +\frac{1}{\sqrt{3}}i\right)
\end{equation}
for the case in hand (see  \cite{Breen:2010ux} for the general form),
$K_{0}$ is a modified Bessel function of the second kind and
\begin{eqnarray}
\xi (r) & = & \left(\int_{1}^{r} \frac{1}{\sqrt{r^2 f(r)}}\right)^{2}
\nonumber \\
& = & \left[ \frac{4}{(D-3)^2} \ln \left(r^{(D-3)/2} +\sqrt{r^{D-3}-1} \right) \right]^{2} . \qquad
\end{eqnarray}
In \cite{Breen:2010ux} it was shown that, for a general $f(r)$ (under the constraints mentioned previously) the approximation (\ref{eq:EGL}) encapsulates enough of the near horizon behaviour of $q_{0\ell}(r)$ in order to capture all of the local contribution (i.e.~the horizon divergence and local finite terms) to the mode sum in (\ref{eq:horizonfinal}) in the limit $r \to 1$. In other words, the sum of the remainder term $\mathcal{R}_{\ell}(r)$ does not contribute in this limit. The $\beta_\ell$ terms in (\ref{eq:qapprox}) are determined by the requirement that $q_{0\ell}(r)$ should vanish as $r \to \infty$.

In the limit $r \to 1$ we then have
\begin{eqnarray}
& & \frac{1}{\sqrt{2\pi T}} \sum_{\ell=0}^{\infty}\left(2 \ell+1 \right)q_{0\ell}(r)
\nonumber \\ & &
= \frac{1}{2\pi T}\left[\sqrt{2\pi T}\sum_{\ell=0}^{\infty}\left( 2\ell+1 \right) Q_{0\ell}(r)
+\sum_{\ell=0}^{\infty}\left( 2\ell+ 1 \right )\beta_{\ell} \right]
\nonumber \\ & &
\qquad +O\left[(r-1) \ln(r-1)\right] .
\label{eq:sum1}
\end{eqnarray}
Therefore in order to calculate $\langle \hat{\phi }^2\rangle_{\text{ren}}$ for the space-time (\ref{eq:Emetric}), we are required to calculate the two summations on the right hand side of (\ref{eq:sum1}) and combine them with the renormalization subtraction terms, before taking the $r \to 1$ limit, namely
\begin{eqnarray}
\langle \hat{\phi }^2\rangle_{\text{ren}} &= &
\lim_{r \to 1}\left[{\sqrt{\frac {T}{32\pi^3}}}\sum_{\ell=0}^{\infty}\left( 2\ell+1 \right) Q_{0\ell}
-\frac{D-3}{16 \pi^2 (r-1)}
\right. \nonumber \\ & & \left.
 +\frac{D-3}{48\pi^2 }\right] +\frac{1}{8\pi^2}\sum_{\ell=0}^{\infty}\left( 2\ell+1 \right) \beta_{\ell} .
\label{eq:phirenf}
\end{eqnarray}
\begin{widetext}
A closed form expression for the first term on the right hand side of (\ref{eq:phirenf}) has been calculated in  \cite{Breen:2010ux}, which here reduces to
\begin{eqnarray}
\frac{1}{96 \pi ^2}\left\{  2 D-5-2 i
   \sqrt{3} \ln \left[\frac{\Gamma \left(a\right)}{\Gamma \left(\bar{a}\right)}\right]
+12 \frac{d}{dx} \zeta\left(x,a\right) \bigg |_{x=-1}
+12 \-\frac{d}{dx} \zeta\left(x,\bar{a}\right) \bigg |_{x=-1}  \right\} ,
\nonumber \\
\label{eq:anal}
\end{eqnarray}
where $\zeta(x,a)$ is the generalised Riemann zeta function.

All that remains now is to calculate the $\beta_{\ell}$ summation in (\ref{eq:phirenf}). In general, this must be performed numerically (indeed in general $\beta_{\ell}$ itself must be obtained numerically for each $\ell$), however in this case, primarily because we have an expression (\ref{eq:qsol}) for $q_{0\ell}(r)$ in terms of hypergeometric functions, we may calculate this sum analytically to give a completely closed form expression for
$\langle {\hat {\phi }}^{2} \rangle _{\text {ren}}$ on the brane for an arbitrary number of bulk space-time dimensions.

Our first step in this process is to find the $\beta_{\ell}$ themselves. We compare the lowest order terms in the expansions of the two expressions (\ref{eq:qsol}) and (\ref{eq:qapprox}) for $q_{0\ell}(r)$. Expanding (\ref{eq:qsol}) gives:
\begin{equation}
q_{0\ell}(r)   =  -\frac{1}{2\sqrt{2\pi T}} \left\{ 2\gamma +\ln\left[ (D-3)(r-1) \right]
+\psi\left( \frac{\ell+D-3}{D-3} +\alpha\right) +\psi\left(\frac{\ell+1}{D-3} -\alpha\right) \right\}
+O\left[ (r-1) \ln(r-1)\right] ,
\label{eq:expan1}
\end{equation}
where $\gamma $ is Euler's constant and $\psi $ is the digamma function, while by expanding (\ref{eq:qapprox}) we obtain:
\begin{equation}
q_{0\ell}(r)  = -\frac{1}{2\sqrt{2\pi T}} \left\{ 2\gamma +\ln\left(\frac{r-1}{D-3}\right)
+ \ln\left[ (\ell+a)(\ell +\bar{a})\right] \right\} +\frac{1}{\sqrt{2\pi T}} \beta_{\ell}
+O\left[(r-1) \ln(r-1)\right].
\label{eq:expan2}
\end{equation}
Comparing the two expansions (\ref{eq:expan1}, \ref{eq:expan2}), we arrive at an expression for $\beta_{\ell} $:
\begin{equation}
\beta_{\ell} = \frac{1}{2}\left\{ \ln\left[ (\ell+a)(\ell +\bar{a} )\right] -2\ln (D-3)
 -\psi\left(\frac{\ell+D-3}{D-3}+\alpha\right) -\psi\left(\frac{\ell+1}{D-3} -\alpha\right)\right\} ,
\label{eq:betaL}
\end{equation}
which, it is straightforward to show, is $O(\ell^{-4}$) as $\ell \to \infty$. Therefore the sum
$\sum_{\ell=0}^{\infty}\left(2 \ell+1 \right)\beta_{\ell}$
is convergent, and as it turns out, amenable to direct calculation. This is  a somewhat involved and technical calculation, with details given in the Appendix.  The result of this analysis is:
\begin{eqnarray}
\frac{1}{8\pi^2}\sum_{\ell=0}^{\infty}\left( 2\ell+1 \right)\beta_{\ell } &  = &
 \frac{1}{96\pi^2} \bigg\{  -1+2 i
 {\sqrt{3}} \ln \left[ \frac{\Gamma \left(a\right)}{\Gamma \left(\bar{a})\right)}\right]
    -12 \frac{d}{dx} \zeta\left(x,a\right) \bigg |_{x=-1}
   -12 \-\frac{d}{dx} \zeta\left(x,\bar{a}\right) \bigg |_{x=-1}  \nonumber \\ & &
+ \frac{6}{D-3}\sum_{j=0}^{D-4} j(j-D+4) \left[ \psi\left(\frac{j+D-3}{(D-3)}+\alpha\right)+\psi\left(\frac{j+1}{(D-3)}-\alpha\right)\right]
\nonumber \\ & &
 +(D-4)(D-5)\left\{\psi\left[(D-3)\left(\alpha+1\right)\right] +\psi\left[1-(D-3)\alpha\right] -2\ln(D-3)\right\}\bigg\}.
\label{eq:apenresult}
\end{eqnarray}
This result has been confirmed numerically.

By combining (\ref{eq:apenresult}) with (\ref{eq:phirenf}, \ref{eq:anal}), and inserting the expression for $\alpha$ (\ref{eq:alpha}) we finally arrive at a closed form expression for the renormalized vacuum polarization $\langle \hat{\phi }^2\rangle_{\text{ren}}$ of a massless conformally coupled scalar field, valid on the horizon of the brane black hole
\begin{eqnarray}
\langle \hat{\phi }^2\rangle_{\text{ren}} & = &
\frac{D-3}{48\pi^2}
\nonumber \\ & &
+ \frac {(D-4)(D-5)}{96\pi ^{2}}\left[ \psi\left(\frac{D-2}{2} -\frac {{\sqrt{(D-4)(D-2)}}}{2{\sqrt {3}}}\right)
+\psi\left(\frac{D-2}{2} +\frac {{\sqrt {(D-4)(D-2)}}}{2{\sqrt {3}}}\right)
-2\ln(D-3) \right]
 \nonumber \\ & &
+ \frac{1}{16\pi ^{2}(D-3)} \sum_{j=0}^{D-4} j(j-D+4) \left[ \psi\left(\frac{6j +3(D-2)-\sqrt{3(D-4)(D-2)}}{6(D-3)}\right)
\right.  \nonumber \\ & & \qquad \left.
+\psi\left(\frac{ 6j +3(D-2) +\sqrt{3(D-4)(D-2)}}{6(D-3)}\right)\right] .
\label{eq:horizonresult}
\end{eqnarray}
\end{widetext}
We analyze this result, together with the renormalized vacuum polarization away from the horizon, in the next section.

\section{Results}
\label{sec:results}

In the previous section, we derived an exact, closed-form expression (\ref{eq:horizonresult}) for the renormalized vacuum polarization on the event horizon of the brane black hole whose metric is given by (\ref{eq:brane}). Outside the event horizon, the renormalized vacuum polarization is given as a sum of two terms, the first of which (\ref{eq:analytic}) is given in closed form, while the second (\ref{eq:numericfinal}) requires numerical computation.  In this section we shall discuss our results for $\langle {\hat {\phi}}^{2} \rangle _{\text {ren}}$ both on and outside the horizon.

\subsection{On the horizon}
\label{sec:horizonresults}

We begin our discussion by examining the expression (\ref{eq:horizonresult}) for $\langle {\hat {\phi}}^{2} \rangle _{\text {ren}}$ on the event horizon.
We see that for $D=4$ and $D=5$, the vacuum polarization on the horizon has the simple form
\begin{equation}
\langle \hat{\phi }^2\rangle_{{\text {ren}}}=\frac{D-3}{48\pi^2},
\label{eq:d1d2}
\end{equation}
which is in agreement with Candelas' result for a Schwarzschild black hole space-time with $D=4$ \cite{Candelas:1980zt}.
Indeed, we may extend Candelas' method to $D=5$ by noting that,  for this case, (\ref{eq:qsol}) reduces to ($2\pi T=1$ (\ref{eq:temp}) for $D=5$ with $r_{h}=1$) \cite{DLMF}:
\begin{equation}
p_{0\ell}(r)= P_{\ell}(r),
\qquad
q_{0\ell}(r) =  Q_{\ell}(r) ,
\end{equation}
where $P_{\ell }$ and $Q_{\ell }$ are Legendre functions.
An application of Heine's formula \cite{DLMF} then allows us to calculate (\ref{eq:horizonresult}), giving, for $D=5$,
\begin{equation}
\langle \hat{\phi }^2\rangle_{\text{ren}}=\frac{1}{24\pi^2} ,
\end{equation}
which is in agreement with (\ref{eq:d1d2}).

\begin{figure}
\includegraphics[width=0.95\columnwidth]{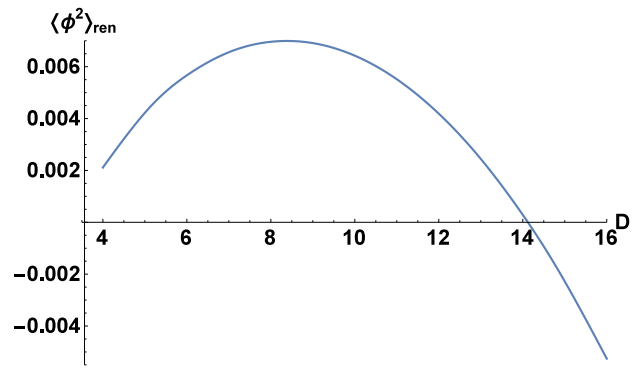}
\caption{Renormalized vacuum polarization $\langle {\hat {\phi }}^{2} \rangle _{\text {ren}}$ on the horizon $r_{h}=1$ of a brane black hole for total number of space-time dimensions $D=4,\ldots, 16$.}
\label{fig:phihor}
\end{figure}

The on-horizon values of $\langle \hat{\phi}^2\rangle_{\text{ren}}$ for $D=4,\ldots , 11$ are given in Tab.~\ref{tab:phihor} and the values for $D=4,\ldots ,16$ are plotted in Fig.~\ref{fig:phihor}. For $D\geq 6$ we give the numerical values of $\langle\hat{\phi}^2\rangle_{\text{ren}}$ in Tab.~\ref{tab:phihor} as they are more informative than the rather unwieldy analytic forms (\ref{eq:horizonresult}).

\begin{table}
\begin{tabular}{|c|c|}
\hline
$D$&$\langle \hat{\phi}^2\rangle_{\text{ren}}$ at $r=1$\\
\hline
4&$1/(48 \pi ^2)$\\
5&$1/(24 \pi ^2)$\\
6&0.0056676345\\
7&0.0065578987\\
8&0.0069605128\\
9&0.0069102842\\
10&0.0064254379\\
11&0.0055159981\\
\hline
\end{tabular}
\caption{Values of the renormalized vacuum polarization $\langle \hat{\phi}^2\rangle_{\text{ren}}$ on the horizon $r_{h}=1$ of a brane black hole for total number of space-time dimensions $D=4,\ldots , 11$.}
\label{tab:phihor}
\end{table}

From inspecting Tab.~\ref{tab:phihor} and Fig.~\ref{fig:phihor} we see that $\langle \hat{\phi}^2\rangle _{\text{ren}}$ reaches its maximum value
on the horizon when the number of bulk space-time dimensions $D=8$, it then decreases and first becomes negative at $D=15$ (note it is still positive when $D=14$). In fact we can see from direct consideration of (\ref{eq:horizonresult}) that the value of $\langle \hat{\phi}^2\rangle _{\text {ren}}$ on the horizon continues to decrease toward $- \infty$ as $D \rightarrow \infty$. We may conclude from this that the magnitude of quantum effects near the horizon grows as the number of bulk space-time dimensions increases. We are working in units in which the event horizon radius $r_{h}$ is fixed to be equal to unity. In these units, the temperature (\ref{eq:temp}) of the brane black hole increases linearly with the number of space-time dimensions $D$.  Therefore, as $D\rightarrow \infty $ with the event horizon radius $r_{h}$ fixed, the temperature of the black hole increases without bound and therefore the vacuum polarization also increases in magnitude. One can therefore argue that as the number of extra bulk dimensions increases, the semi-classical approximation breaks down, at least in the vicinity of the black hole horizon.
When we say that the semi-classical approximation breaks down, we mean that quantum effects are not small when $D$ is large and the back-reaction of the quantum field on the space-time geometry can no longer be ignored. Eventually a full theory of quantum gravity would be required to gain meaningful physical insights.
In an upcoming paper \cite{Breen2015}, it will be shown that the same conclusion may be drawn for the vacuum polarization on the horizon in the bulk of the Schwarzschild-Tangherlini black-hole space-time (\ref{eq:bulk}).

\subsection{$\langle {\hat {\phi }}^{2} \rangle _{\text {analytic}}$}
\label{sec:analytic}

Next we turn to the renormalized vacuum polarization away from the horizon.
The analytic part $\langle {\hat {\phi }}^{2} \rangle _{\text {analytic}}$ (\ref{eq:analytic}), calculated using the metric function (\ref{eq:metricf}),
takes the form
\begin{equation}
\langle {\hat {\phi }}^{2} \rangle _{\text {analytic}} =
\frac {D-3}{192 \pi ^{2}f} \left[ D-3 +
\frac {8 -2D}{r^{D-1}}+\frac {D-5}{r^{2D-4}}
\right] .
\label{eq:analD}
\end{equation}
At the horizon $r=1$ the expression (\ref{eq:analD}) simplifies to
\begin{equation}
\left. \langle {\hat {\phi }}^{2} \rangle _{\text {analytic}} \right| _{r=1} =
\frac {D-3}{48\pi ^{2}}.
\label{eq:analDhor}
\end{equation}
This matches the first term in the exact expression for the renormalized vacuum polarization on the horizon (\ref{eq:horizonresult}).
Furthermore, this is exactly $\langle {\hat {\phi }}^{2} \rangle _{\text {ren}}$ on the horizon when $D=4$ or $5$ (\ref{eq:d1d2}).
The remainder of (\ref{eq:horizonresult}) (which is nonzero only for $D>5$) will come from the numeric part
$\langle {\hat {\phi }}^{2} \rangle _{\text {numeric}}$.
The expression (\ref{eq:analDhor}) is positive for all $D\ge 4$, and linearly increasing as the number of bulk dimensions increases.
As $r\rightarrow \infty $, the expression (\ref{eq:analD}) simplifies to
\begin{equation}
\left. \langle {\hat {\phi }}^{2} \rangle _{\text {analytic}} \right| _{r\rightarrow \infty } =
\frac {(D-3)^{2}}{192\pi ^{2}} = \frac {T^{2}}{12},
\label{eq:analDinf}
\end{equation}
where $T$ is the Hawking temperature (\ref{eq:temp}).
The asymptotic form (\ref{eq:analDinf}) is simply the vacuum polarization for a quantum scalar field at temperature $T$ in flat space-time.
Like the renormalized vacuum polarization (\ref{eq:analDhor}) at the horizon, the form (\ref{eq:analDinf}) at infinity is also positive, but it increases quadratically as the number of bulk dimensions increases.

\begin{figure}
\includegraphics[width=0.95\columnwidth]{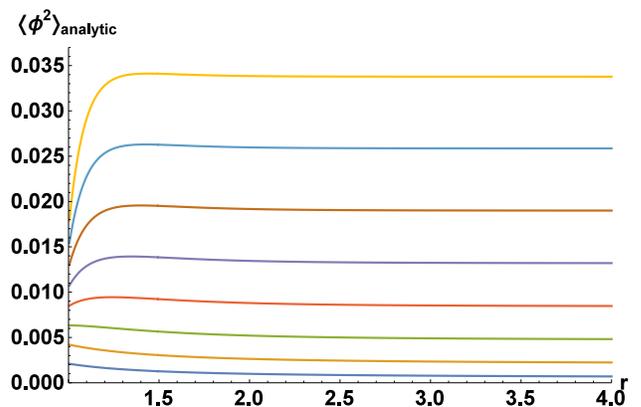}
\caption{Analytic contribution $\langle {\hat {\phi }}^{2} \rangle _{\text {analytic}}$ to the renormalized vacuum polarization on a brane black hole as a function of the radial co-ordinate $r$. The event horizon is located at $r=1$.  The curves (from bottom to top) are for total number of space-time
dimensions $D=4,\ldots , 11$.}
\label{fig:analytic}
\end{figure}

In Fig.~\ref{fig:analytic} we plot $\langle {\hat {\phi }}^{2} \rangle _{\text {analytic}}$ (\ref{eq:analD}) as a function of the radial co-ordinate $r$
for total number of space-time dimensions $D=4,\ldots ,11$ (from bottom to top curves in the figure).
It can be seen that $\langle {\hat {\phi }}^{2} \rangle _{\text {analytic}}$ is positive everywhere on and outside the event horizon, for all $D$.  For
$D=4, 5, 6$ the maximum of $\langle {\hat {\phi }}^{2} \rangle _{\text {analytic}}$ is on the horizon, and it is monotonically decreasing as $r$ increases.
For $D\ge 7$ the maximum of $\langle {\hat {\phi }}^{2} \rangle _{\text {analytic}}$ is a little outside the horizon.

In \cite{Candelas:1984pg} it was found that $\langle {\hat {\phi }}^{2} \rangle _{\text {analytic}}$ was the dominant contribution to the total
renormalized vacuum polarization $\langle {\hat {\phi }}^{2} \rangle _{\text {ren}}$ on a four-dimensional Schwarzschild black hole.
To see if this remains true on the brane when we have extra bulk dimensions, we now examine
the numerical contribution $\langle {\hat {\phi }}^{2} \rangle _{\text {numeric}}$.

\subsection{$\langle {\hat {\phi }}^{2} \rangle _{\text {numeric}}$}
\label{sec:numeric}

The first step in calculating $\langle {\hat {\phi }}^{2} \rangle _{\text {numeric}}$ is to find the mode functions $p_{\omega \ell }(r)$ and
$q_{\omega \ell }(r)$ by numerically integrating the radial equation (\ref{eq:radial}).
The radial equation (\ref{eq:radial}) has a regular singular point at the event horizon $r=1$ and an irregular singular point as $r\rightarrow \infty $.

To find the mode functions $p_{\omega \ell }(r)$, which are regular at the horizon, we start our integration close to $r=1$, using the power series
\begin{equation}
p_{\omega \ell } (r) = \sum _{j=0}^{\infty } a_{j} \left( r- 1 \right) ^{\nu + j},
\end{equation}
where $\nu = \omega /(D-3)$ and $a_{0}$ is set equal to unity, to give suitable initial values for $p_{\omega \ell }(r)$ and its derivative at the starting point.

The modes $q_{\omega \ell }(r)$ are regular as $r\rightarrow \infty $ and have the following asymptotic series near infinity
\begin{equation}
q_{\omega \ell } (r) \sim e^{-\omega r} \sum _{j=0}^{\infty } b_{j} r^{-\varsigma - j},
\end{equation}
where
\begin{equation}
\varsigma =
\begin{cases}
1+ \omega, & \qquad D= 4, \\
1, & \qquad D\ge 5,
\end{cases}
\end{equation}
and we can set $b_{0}=1$ without loss of generality.
To compute $q_{\omega \ell }(r)$ numerically, we found it easier to define a new function ${\tilde {q}}_{\omega \ell }(r)=e^{\omega r} q_{\omega \ell }(r)$
and numerically solve the resulting differential equation for ${\tilde {q}}_{\omega \ell} (r)$, integrating from large $r$ down to near the horizon.

Since the mode sum in (\ref{eq:numericfinal}) involves cancellations between the product of the mode functions $p_{\omega \ell }(r)$, $q_{\omega \ell }(r)$
and the WKB expansion terms $\zeta _{k\omega \ell }(r)$, we require the mode functions to a high degree of precision. All numerical calculations were performed in {\texttt{Mathematica}}, in which computations at the required precision are straightforward to implement. The accuracy of our numerical integration was checked by evaluating the normalization constant $C_{\omega \ell }$ using (\ref{eq:C}) at each value of $r$ in our integration grid.
For all mode functions calculated, $C_{\omega \ell }$ remained constant in $r$ up to 26 significant figures.

Once the mode functions have been found, the mode sum in (\ref{eq:numericfinal}) is straightforward to compute as the WKB terms $\zeta _{k\omega \ell }(r)$, while complicated, are algebraic expressions.  The numerical integrals $J_{k}$ (\ref{eq:Jint2}) are computed using standard numerical integration routines in {\texttt{Mathematica}}.

\begin{figure}
\includegraphics[width=0.95\columnwidth]{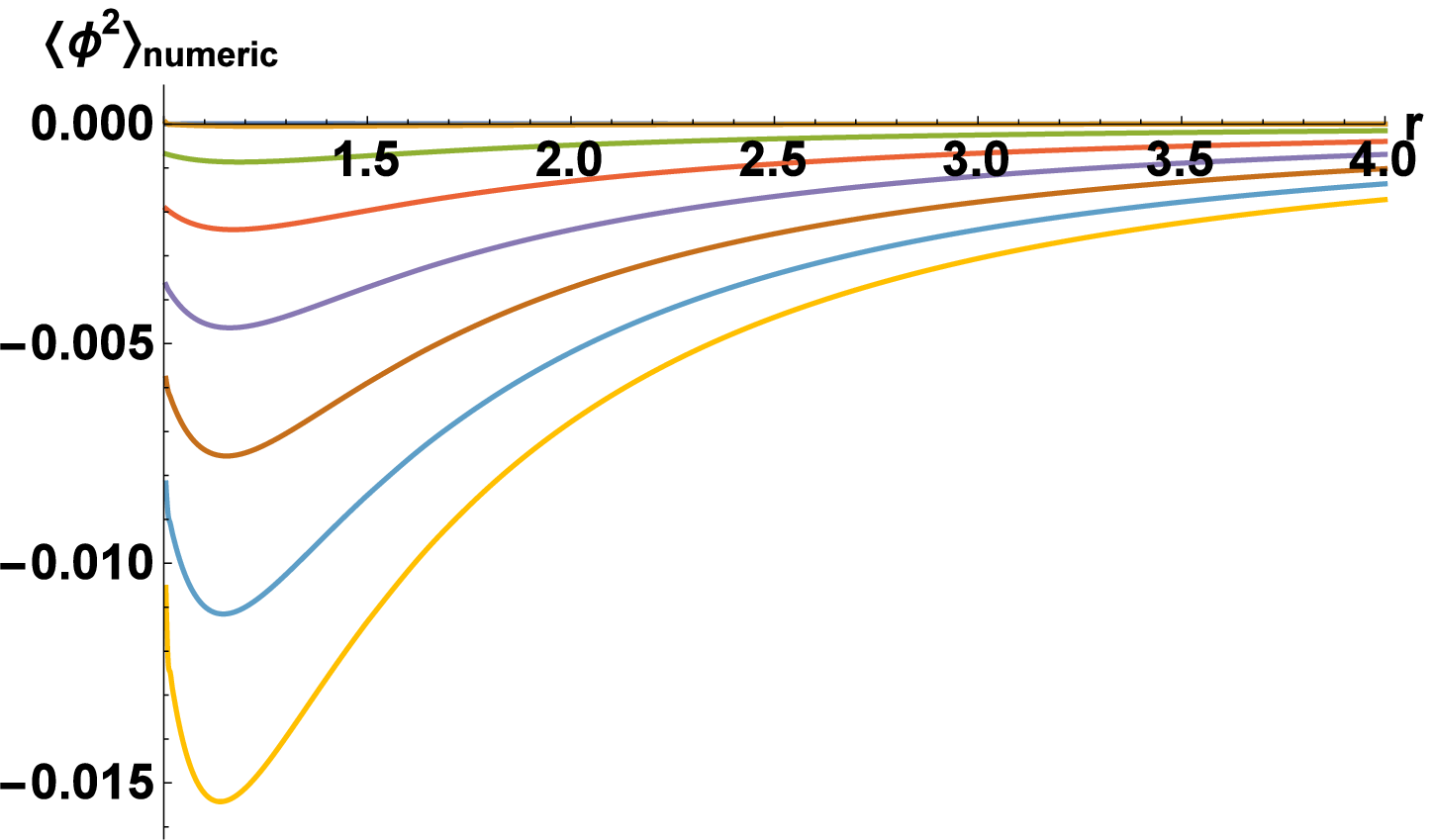}
\medskip
\newline
\includegraphics[width=0.95\columnwidth]{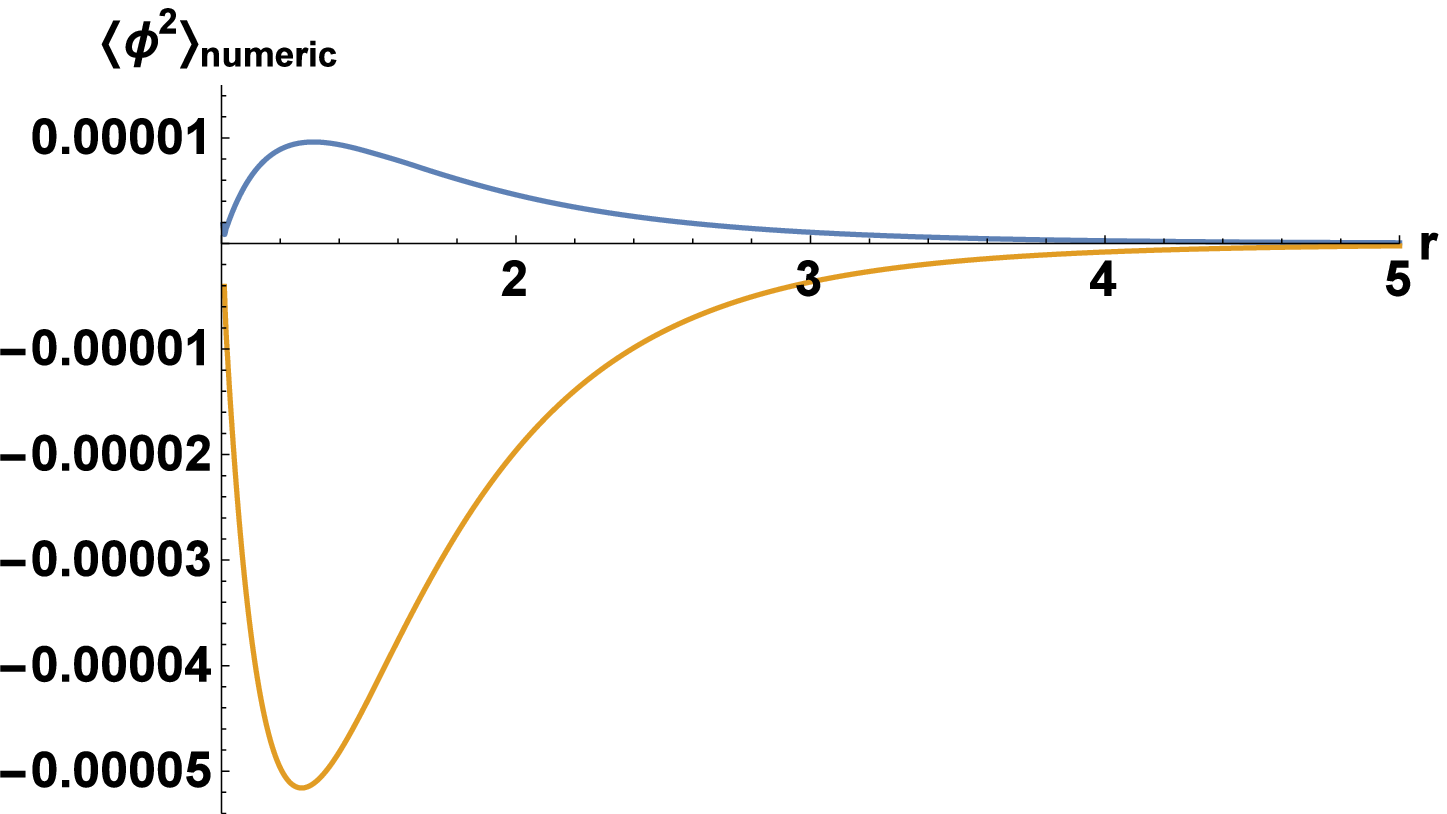}
\caption{Numeric contribution $\langle {\hat {\phi }}^{2} \rangle _{\text {numeric}}$ to the renormalized vacuum polarization on a brane black hole as a function of the radial co-ordinate $r$. The event horizon is located at $r=1$.
(Top plot) The curves (from top to bottom) are for total number of space-time
dimensions $D=4,\ldots , 11$.
(Bottom plot) The curves (from top to bottom) are for $D=4, 5$ respectively.}
\label{fig:numeric}
\end{figure}

Our results for $\langle {\hat {\phi }}^{2} \rangle _{\text {numeric}}$ on the brane as a function of the radial co-ordinate $r$ are shown in Fig.~\ref{fig:numeric} for total number of space-time dimensions $D=4,\ldots ,11$.
In the top plot in Fig.~\ref{fig:numeric} we show all eight curves for $D=4,\ldots ,11$ (from top to bottom).  In this plot, the curves for $D=4,5$ are virtually superimposed on the horizontal axis.
To see these more clearly, in the bottom plot in Fig.~\ref{fig:numeric} we show just the curves for $D=4,5$.

For $D=4$, in the bottom plot in Fig.~\ref{fig:numeric} $\langle {\hat {\phi }}^{2} \rangle _{\text {numeric}}$ is positive everywhere outside the event horizon.
We have compared our numerical values for $\langle {\hat {\phi }}^{2} \rangle _{\text {numeric}}$ with those tabulated in \cite{Candelas:1984pg}
and the agreement is excellent.
For $D=4$, it can be shown that the numeric contribution $\langle {\hat {\phi }}^{2} \rangle _{\text {numeric}}$ vanishes on the event horizon $r=1$.
From Fig.~\ref{fig:numeric} it has a maximum just outside the event horizon and decays to zero quickly as $r$ increases.
Comparing Figs.~\ref{fig:analytic} and \ref{fig:numeric}, for $D=4$ the numeric contribution $\langle {\hat {\phi }}^{2} \rangle _{\text {numeric}}$
to the total renormalized vacuum polarization $\langle {\hat {\phi }}^{2} \rangle _{\text {ren}}$ is negligible compared to the
analytic contribution $\langle {\hat {\phi }}^{2} \rangle _{\text {analytic}}$ for all $r$, again in agreement with \cite{Candelas:1984pg}.

For $D\ge 5$, the numeric contribution $\langle {\hat {\phi }}^{2} \rangle _{\text {numeric}}$ is negative for all $r>1$.
When $D=5$, comparing Figs.~\ref{fig:analytic} and \ref{fig:numeric} again reveals that $\langle {\hat {\phi }}^{2} \rangle _{\text {numeric}}$ is
negligible compared to $\langle {\hat {\phi }}^{2} \rangle _{\text {analytic}}$, but this is not the case for $D\ge 6$.
For all $D\ge 5$ we find that $\langle {\hat {\phi }}^{2} \rangle _{\text {numeric}}$ has a minimum just outside the event horizon, and increases toward zero
as $r$ increases.  We also find that the magnitude of $\langle {\hat {\phi }}^{2} \rangle _{\text {numeric}}$ for fixed $r$ is increasing as the number
of bulk space-time dimensions $D$ increases.

\subsection{Total renormalized vacuum polarization}
\label{sec:total}

\begin{figure}
\includegraphics[width=0.95\columnwidth]{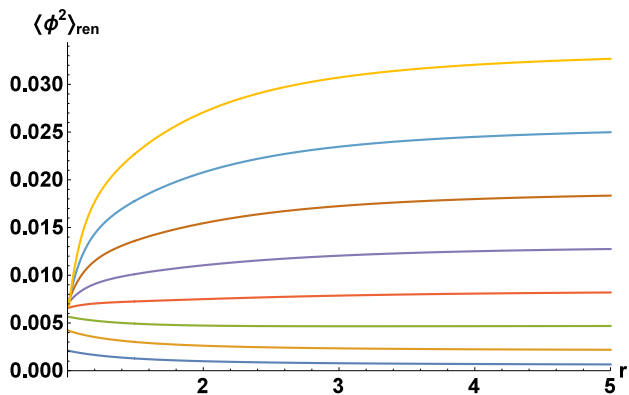}
\caption{Total renormalized vacuum polarization $\langle {\hat {\phi }}^{2} \rangle _{\text {ren}}$ on a brane black hole as a function of the radial co-ordinate $r$. The event horizon is located at $r=1$.  The curves (from bottom to top) are for total number of space-time
dimensions $D=4,\ldots , 11$.}
\label{fig:total}
\end{figure}

We now combine the results of the previous two subsections to find the total renormalized vacuum polarization
$\langle {\hat {\phi }}^{2} \rangle _{\text {ren}}$ on the brane for a higher-dimensional Schwarzschild-Tangherlini black hole.
Our results are presented in Fig.~\ref{fig:total}, where we plot $\langle {\hat {\phi }}^{2} \rangle _{\text {ren}}$  as a function of the radial
co-ordinate $r$ for (from bottom to top curves) total number of space-time dimensions $D=4,\ldots , 11$.
Our numerical calculations for $\langle {\hat {\phi }}^{2} \rangle _{\text {numeric}}$ are valid only outside the horizon, for $r>1$. We have extrapolated
our numerical results for $\langle {\hat {\phi }}^{2} \rangle _{\text {ren}}$ for $r>1$ to the horizon at $r=1$ and find excellent agreement with the exact results
on the horizon given in Tab.~\ref{tab:phihor}.

For $D=4,\ldots, 11$ we see from Fig.~\ref{fig:total} that the renormalized vacuum polarization $\langle {\hat {\phi }}^{2} \rangle _{\text {ren}}$
is positive everywhere on and outside the event horizon.
For $D=4,5,6$ it has its maximum value on the horizon, and is monotonically decreasing as $r$ increases.
For $D\ge 7$ we find that  $\langle {\hat {\phi }}^{2} \rangle _{\text {ren}}$  has a minimum value on the horizon and is monotonically increasing as $r$ increases and we move away from the event horizon.
For all $D$, the numeric contribution $\langle {\hat {\phi }}^{2} \rangle _{\text {numeric}}$ becomes insignificant for large $r$, and the analytic
contribution $\langle {\hat {\phi }}^{2} \rangle _{\text {analytic}}$ dominates.
However, close to the horizon the numeric part $\langle {\hat {\phi }}^{2} \rangle _{\text {numeric}}$ makes a significant contribution to the total.

Far from the event horizon of the brane black hole, the vacuum polarization $\langle {\hat {\phi }}^{2} \rangle _{\text {ren}}$ increases rapidly as the
number of bulk space-time dimensions $D$ increases.
As explained earlier, this is to be expected because we are working in units in which the event horizon radius $r_{h}$ is fixed to be unity, and in
these units the temperature of the brane black hole (\ref{eq:temp}) increases linearly with the number of bulk space-time dimensions $D$.

Close to the horizon, due to the significant negative contribution to the total from the numeric part $\langle {\hat {\phi }}^{2} \rangle _{\text {numeric}}$, the vacuum polarization is rather smaller than it is far from the black hole. As shown in Fig.~\ref{fig:phihor}, for $D=4,\ldots, 11$ the vacuum polarization is always positive on the horizon and has a maximum when $D=8$.

\section{Conclusions}
\label{sec:conc}

In this paper we have calculated the renormalized vacuum polarization $\langle {\hat {\phi }}^{2} \rangle _{\text {ren}}$ for a massless, conformally coupled, quantum scalar field ${\hat {\phi }}$ propagating
on the brane of a higher-dimensional Schwarzschild-Tangherlini black hole.
Our four-dimensional, on-brane metric, is static and spherically symmetric and corresponds to a slice of the higher dimensional black hole geometry. The metric functions depend on the total number of space-time dimensions $D$, and in this paper we focus on $D=4,\ldots , 11$, in common with the literature on Hawking radiation from brane black holes
\cite{Kanti:2004nr,Park:2012fe,Casanova:2005id,Kanti:2008eq,Kanti:2012jh,Winstanley:2007hj,Kanti:2014vsa}.
The Hawking radiation of a scalar field from such a brane black hole has been studied in depth (see, for example, \cite{Harris:2003eg})
but the Hawking fluxes can be computed without recourse to renormalization.
To the best of our knowledge, we have presented in this paper the first computation of an expectation value requiring renormalization on a brane black hole space-time.

Since the on-brane metric is four-dimensional, static and spherically symmetric, we have been able to employ well-established methodology \cite{Anderson:1993if,Anderson:1994hg,Winstanley:2007tf,Breen:2010ux} to compute the renormalized vacuum polarization.
We work on Euclidean space-time and consider the scalar field to be in the Hartle-Hawking state \cite{Hartle:1976tp}, a thermal quantum state at the Hawking temperature $T$.
We use covariant geodesic point separation to regularize the Euclidean Green's function and then the divergent subtraction terms are given by the Hadamard form.
Away from the horizon, the space-time points are split in the temporal direction.  In this case the renormalized vacuum polarization can be written
as the sum of two terms, the first of which is a simple closed-form expression, while the second requires numerical computation.
On the horizon, radial point-splitting is used and an exact (but complicated) closed-form expression for the renormalized vacuum polarization is found.

For $D=4,\ldots, 11$ we find that the renormalized vacuum polarization is positive everywhere on and outside the event horizon. Far away from the horizon, its value increases rapidly as the number of bulk space-time dimensions $D$ increases. This is because the on-brane temperature of the black hole increases linearly with increasing $D$ for fixed event horizon radius $r_{h}$ and we are considering a thermal quantum state.
Ultimately, for $D$ very large (with fixed $r_{h}$) the semi-classical approximation used here breaks down in the sense that quantum effects are no longer small and the back-reaction of the quantum field on the black hole
geometry can no longer be ignored to first order.  This is also the case for very small black holes which have entered the Planck phase of their evolution \cite{Giddings:2001bu}.  In either of these regimes ($D$ very large or very small black holes) a full theory of quantum gravity is required to model the behaviour of the black holes.
Close to the horizon, the renormalized vacuum polarization becomes negative when $D\ge 15$ and decreases rapidly as $D$ increases further.  This again
ultimately leads to a breakdown in the semi-classical approximation when $D$ is very large.

In this paper we have taken the quantum scalar field to be in the Hartle-Hawking state \cite{Hartle:1976tp} since this state is the easiest in which
to compute renormalized expectation values.  The quantum state of interest for simulations of brane black holes at the LHC \cite{Dai:2007ki,Frost:2009cf}
is the Unruh state \cite{Unruh:1976db} since this state represents an evaporating black hole. Since differences in expectation values between two quantum
states do not require renormalization, it would be comparatively straightforward to compute the renormalized vacuum polarization in the Unruh state using the results presented here for the Hartle-Hawking state.

We have considered the renormalized vacuum polarization $\langle {\hat {\phi }}^{2} \rangle _{\text {ren}}$ because it is the simplest nontrivial expectation value for a quantum scalar field. While it is a scalar (and hence cannot, for example, distinguish between the future and past event horizons of a black hole), it nonetheless shares some physical features with the renormalized stress-energy tensor $\langle {\hat {T}}_{\mu \nu} \rangle _{\text {ren}}$.
The renormalized stress-energy tensor is of particular interest because it governs the back-reaction of the quantum field on the space-time geometry via the semi-classical Einstein equations (\ref{eq:scee}).
However, since $\langle {\hat {T}}_{\mu \nu} \rangle _{\text {ren}}$ is a tensor object and involves derivatives of the Green's function, it is more complicated to calculate than $\langle {\hat {\phi }}^{2} \rangle _{\text {ren}}$, although the standard methodology \cite{Anderson:1994hg,Breen:2011aa} should be applicable for the on-brane metric considered in this paper.
It would be interesting to investigate whether the effects we have found here for the renormalized vacuum polarization (increasing magnitude with increasing bulk dimension $D$ and negative values near the horizon for large $D$) are present also in $\langle {\hat {T}}_{\mu \nu} \rangle _{\text {ren}}$.

There are other extensions to this work on the brane which would be of interest. In this paper we have considered a massless, conformally coupled scalar field. It is known that including a mass suppresses the Hawking radiation on the brane and so the emission of light quantum fields is of the greatest interest phenomenologically. For a four-dimensional Schwarzschild black hole, the Hawking radiation is dominated by scalar fields but as the number of bulk dimensions
increases, the on-brane emission of particles with non-zero spin becomes of comparable magnitude to scalar emission \cite{Harris:2003eg}.
Emission of spin-half quanta has particular phenomenological importance given the large number of fermion degrees of freedom in the Standard Model.
Therefore the computation of $\langle {\hat {T}}_{\mu \nu} \rangle _{\text {ren}}$ for non-zero spin fields, while technically challenging, would be of interest.

Here we have restricted our attention to static, nonrotating black holes.  Hawking radiation on the brane from rotating black holes has been extensively studied (see, for example, \cite{Kanti:2014vsa,Winstanley:2007hj} for reviews) but the computation of renormalized expectation values everywhere outside the horizon of a four-dimensional rotating black hole has proved intractable to date.

Finally we emphasize that our results in this paper are for the renormalized vacuum polarization on the four-dimensional brane metric of a higher-dimensional black hole.  For smaller values of $D$, emission on the brane dominates that in the bulk \cite{Casals:2008pq,Emparan:2000rs,Harris:2003eg}, but as the number of bulk dimensions $D$ increases, Hawking radiation of gravitons (and possibly scalar fields) in the bulk does become significant.
While the formalism underlying the renormalization of expectation values for quantum scalar fields on higher-dimensional space-times is known \cite{Decanini:2005eg,Thompson:2008bk}, a methodology for detailed computations in the full space-time exterior to the event horizon has yet to
be developed.  We leave these questions for future work.

\bigskip
\bigskip
\bigskip
\bigskip
\appendix
\section{$\beta_{\ell}$ summation}

In this Appendix we give the details of the derivation of (\ref{eq:apenresult}).
We begin with (\ref{eq:betaL})
\begin{eqnarray}
\beta_{\ell} & = &
\frac{1}{2} \ln\left(\ell+a\right)
+\frac{1}{2} \ln\left(\ell+\bar{a}\right) -\ln(D-3)
\nonumber\\
& & -\frac{1}{2}\psi\left(\frac{\ell+D-3}{D-3} +\alpha\right)-\frac{1}{2}\psi\left(\frac{\ell+1}{D-3} -\alpha \right).\nonumber\\
\end{eqnarray}
In order to find the sum $\sum_{\ell =0}^{\infty } \left( 2\ell+ 1 \right) \beta _{\ell }$, we consider instead the sum
\begin{equation}
\Sigma (z)=\sum _{\ell =0}^{\infty} \beta _{\ell }z^{\ell }.
\label{eq:Sigma}
\end{equation}
Our strategy is to expand $\Sigma (z)$ and its derivative about $z=1$, which we may then combine to give
\begin{equation}
 \sum_{\ell=0}^{\infty}\left(2\ell+1\right)\beta_\ell =  2\Sigma '(1)  + \Sigma (1).
\end{equation}
upon taking the limit $z\rightarrow 1$.

We begin by examining the expression
\begin{eqnarray}
& & \sum_{\ell=0}^{\infty}\psi\left(\frac{\ell}{p} +q\right)z^{\ell}
=\frac{1}{1-z^p}\left\{\sum_{\ell=0}^{p-1} \psi\left(\frac{\ell}{p} +q\right)z^{\ell} \right.\nonumber \\
& & \left. \quad + \sum_{\ell=p}^{\infty} \left[ \psi\left(\frac{\ell}{p} +q\right)z^{\ell}-\psi\left(\frac{\ell}{p} +q-1\right)z^{\ell}\right] \right\}.
\end{eqnarray}
Introducing a new variable $k=\ell-p$ and making use of Eq.~(8.365.1) of \cite{GradRiz}
\begin{equation}
\psi(\upsilon+1)-\psi(\upsilon)=\frac{1}{\upsilon},
\end{equation}
we arrive at
\begin{eqnarray}
\sum_{\ell=0}^{\infty}\psi\left(\frac{\ell}{p} +q\right)z^{\ell}
& = &\frac{1}{1-z^p}\left[\sum_{\ell=0}^{p-1} \psi\left(\frac{\ell}{p} +q\right)z^{\ell}\right.\nonumber\\
& &\left.  +p z^p \sum_{k=0}^{\infty} \frac{z^k}{k +pq}\right].\nonumber\\
\label{eqn:psi}
\end{eqnarray}
From Eq.~(8.365.6) in  \cite{GradRiz}, we have
\begin{equation}
\sum_{\ell=0}^{p-1} \psi\left(\frac{\ell}{p}+z\right)=p\left[ \psi(p z)-\ln(p) \right] ,
\end{equation}
and so the right hand side of (\ref{eqn:psi}) can be written as
\begin{eqnarray}
& & \frac{1}{1-z^p}\left[\sum_{\ell=0}^{p-1} \psi\left(\frac{\ell}{p}+q\right)(z^\ell-1)
\right. \nonumber \\ & & \left.
\quad +p\left\{z^p \sum_{\ell=0}^{\infty} \frac{z^k}{k+pq}
+\psi\left(pq\right)-\ln(p)\right\}\right].
\end{eqnarray}
The definition of the hypergeometric function \cite{DLMF} gives,
\begin{eqnarray}
\frac{1}{pq}
{}_2 F_1\left(1 , pq; 1+pq;z\right)&=&\sum_{k=0}^{\infty} \frac{z^k}{k +pq}.
\end{eqnarray}

Combining the above work, we may split our sum $\Sigma (z)$ (\ref{eq:Sigma}) into three separate parts
\begin{equation}
\Sigma (z)=S_1+S_2+S_3,
\end{equation}
where
\begin{widetext}
\begin{eqnarray}
S_1 & = &
\frac {1}{2} \sum_{\ell=0}^{\infty}\left[ \ln(\ell+a)
+ \ln(\ell+\bar{a}) \right] z^\ell ,
\nonumber \\
S_2 & = & -\frac{1}{2\left( 1-z^{D-3} \right) }\sum_{\ell=0}^{D-4}\left[\psi\left(\frac{\ell}{D-3} +\alpha+1\right)
+ \psi\left( \frac{\ell+1}{D-3} -\alpha \right)\right](z^\ell-1) ,
\nonumber \\
S_3 & = & -\frac{D-3}{2\left( 1-z^{D-3} \right)} \left\{
\frac{z^{D-3}}{(D-3)(\alpha+1 )}
{}_2 F_1\left(1, (D-3)(\alpha+1 ); 1+(D-3)(\alpha+1 ); z\right)
+ \psi\left[ (D-3)(\alpha+1 )\right]
\right. \nonumber \\ & & \left.
 +\psi\left[ 1-(D-3)\alpha \right]
+\frac{z^{D-3}}{1-(D-3)\alpha}
{}_2 F_1\left(1, 1-(D-3)\alpha ; 2-(D-3)\alpha ; z \right)
\right\} -\frac{\ln(D-3)}{1-z} +\frac{(D-3)\ln(D-3)}{1-z^{D-3}}.
\nonumber \\ & &
\end{eqnarray}
We now proceed to expand about $z=1$, retaining  the $O(z-1)$ terms as we will eventually be taking the derivative of the resulting expressions with respect to $z$.
We will deal with $S_2$ and $S_3$ first as they are the most straightforward to work with. The required expansions are readily obtained using {\texttt{Mathematica}}:
\begin{eqnarray}
S_2 & = &
\frac{1}{2(D-3)}\sum_{\ell=0}^{D-4}\ell\left[\psi\left(\frac{\ell}{D-3} +\alpha+1\right)
+ \psi\left(\frac{\ell+1}{D-3} -\alpha \right)\right]
\nonumber \\ & &
+\frac{z-1}{4(D-3)}\left\{ \sum_{\ell=0}^{D-4}\ell(\ell+3-D)
\left[\psi\left(\frac{\ell}{D-3} +\alpha+1\right)  + \psi\left( \frac{\ell+1}{D-3} -\alpha \right)\right]\right\} +O(z-1)^2,
\label{eq:S2expan}
\end{eqnarray}
and
\begin{eqnarray}
S_3 & = &
\frac{1}{2}\left\{ (D-3)\alpha \psi \left[(D-3)(\alpha+1 )\right] - \left[(D-3)\alpha +D-4\right]
\psi \left[ 1-(D-3)\alpha\right] +(D-4) \left[\ln (D-3)-1\right] \right\}
\nonumber \\ & &
+\frac{z-1}{24}  \left\{
\left[ (D-4)(D-5)-6(D-3)\alpha\right] \psi\left[ (D-3)(\alpha+1)\right]
\right. \nonumber \\ & & \left.
+\left[ (D-4)(D+1)+6 (D-3)\alpha\right] \psi \left[ 1-(D-3)\alpha\right]
+6 (D-4)-2 (D-2)   (D-4) \ln (D-3)\right\}
\nonumber \\ & &
-\frac{\ln  (1-z)+\gamma }{z-1}
+O\left[(z-1)^2\right].
\label{eq:S3expan}
\end{eqnarray}
Returning to $S_1$,  from Eq.~(9.550) in \cite{DLMF} we have that
\begin{equation}
\sum_{\ell=0}^{\infty}\frac{z^\ell}{(\upsilon+\ell)^s}=\Phi(z,s,\upsilon), \qquad  |z|<1 ,
\end{equation}
where $\Phi(z,s,\upsilon)$ is Lerch's transcendent function.
Therefore
\begin{equation}
\sum_{\ell=0}^{\infty}z^{\ell }\ln\left(\ell+\upsilon\right) =\frac{d}{ds}\Phi(z,s,\upsilon)|_{s=0}
\end{equation}
and so we can express $S_1$ in the form
\begin{equation}
S_1=-\frac{1}{2} \left[ \left. \frac{d}{ds}\Phi\left(z,s,a \right)\right|_{s=0}
+ \left. \frac{d}{ds}\Phi\left(z,s,\bar{a}\right)\right|_{s=0}\right].
\label{eq:S1}
\end{equation}
To expand this expression about $z=1$ we make use of the following result for $\Phi(z,s,\upsilon)$ (see Sec.~1.11 in \cite{Erdelyi:1953:HTF2})
\begin{equation}
\Phi(z,s,\upsilon)=\frac{\Gamma(1-s)}{z^\upsilon}\left[ \ln \left( \frac{1}{z}\right)\right]^{s-1}+ z^{-\upsilon}\sum_{k=0}^{\infty}\zeta(s-k,\upsilon) \frac{(\ln z)^k}{k!}.
\end{equation}
Taking derivatives of both sides with respect to $s$ and setting $s=0$ yields
\begin{eqnarray}
\frac{d}{ds}\Phi(z,s,\upsilon)|_{s=0} & = &
\frac{1}{z^\upsilon}\left\{
\ln(\Gamma(\upsilon))-\ln(\sqrt{2\pi}) +\frac{\gamma+\ln(\ln(1/z))}{\ln(1/z)}
+\sum_{k=1}^{\infty}\frac{d}{ds}\zeta(s-k,\upsilon)\bigg|_{s=0} \frac{(\ln z)^k}{k!}\right\} ,
\label{eq:S1alt}
\end{eqnarray}
where we have used the result \cite{DLMF}
\begin{equation}
\left. \frac{d}{ds}\zeta(s,\upsilon)\right|_{s=0}=\ln(\Gamma(\upsilon))-\ln(\sqrt{2\pi}).
\end{equation}
Expanding (\ref{eq:S1alt}) about $z=1$ and inserting the resulting expression into (\ref{eq:S1}) yields:
\begin{eqnarray}
S_1 & = & \frac{1}{2}\left[ \ln(2\pi)-1 -\ln\Gamma(a)-\ln\Gamma(\bar{a}) \right]
+\frac{z-1}{24} \bigg\{ 5-6\ln(2\pi) +12\left[ a \ln \Gamma(a)+\bar{a} \ln \Gamma(\bar{a}) \right]
 \nonumber \\ & &
-12\left( \left. \frac{d}{ds}\zeta(s-1,a)\right|_{s=0}+ \left. \frac{d}{ds}\zeta(s-1,\bar{a})\right|_{s=0}\right)\bigg\}
+\frac{\ln (1-z)+\gamma }{z-1}+O\left[(z-1)^2\right].
\label{eq:S1expan}
\end{eqnarray}
Combining (\ref{eq:S2expan}, \ref{eq:S3expan}, \ref{eq:S1expan}) yields
\begin{equation}
\Sigma (z)= F_1+ F_2(z-1)+ O\left[(z-1)^2\right] ,
\label{eq:betas}
\end{equation}
where
\begin{eqnarray}
F_1 & = &
\frac {1}{2} \bigg\{
\left[\ln\left(\frac{2\pi}{D-3}\right) +(D-3)\left[\ln (D-3)-1\right]-\ln \Gamma( a)-\ln\Gamma( \bar{a})\right]
+\alpha(D-3)\psi\left[(D-3)(\alpha+1)\right]
 \nonumber \\ & &
-\left[ (D-3)\alpha+(D-4)\right] \psi\left[1-(D-3)\alpha\right] \bigg\}
+\frac {1}{2(D-3)}\sum_{\ell=0}^{D-4}\ell\left\{ \psi\left[\frac{\ell}{D-3}+\alpha+1\right]+ \psi\left[\frac{\ell+1}{D-3}-\alpha\right] \right\},
\nonumber \\
F_2 & = &
\frac{1}{24}\bigg\{ 6D-19 +2 \ln(D-3)-6\ln(2\pi) -2(D-3)^2 \ln(D-3)
+12a \ln \Gamma(a) +12\bar{a} \ln \Gamma(\bar{a})
 \nonumber \\ & &
 + \left[ (D-5)(D-4)-6(D-3)\alpha\right]\psi\left[ (D-3)(\alpha+1) \right]
  \nonumber \\ & &
+  \left[ (D-4)(D-5)+6(D-3)\alpha\right] \psi\left[ 1-(D-3)\alpha \right]
-12\left[ \left. \frac{d}{ds}\zeta(s-1,a)\right|_{s=0}+ \left. \frac{d}{ds}\zeta(s-1,\bar{a})\right|_{s=0}\right] \bigg\}
\nonumber \\ & &
+\frac {1}{4(D-3)}\sum_{\ell=0}^{D-4}\ell(\ell+3-D) \left[ \psi\left(\frac{\ell}{D-3}+\alpha+1\right)+ \psi\left(\frac{\ell+1}{D-3}-\alpha\right)\right] .
\end{eqnarray}
\end{widetext}
Finally, by taking the derivative of (\ref{eq:betas}) with respect to $z$ and setting $z=1$ we obtain:
\begin{equation}
\sum_{\ell=0}^{\infty}\left( 2\ell+1 \right )\beta_\ell = 2F_2 +F_1,
\end{equation}
which simplifies to the expression contained in (\ref{eq:apenresult}).

\bigskip

\begin{acknowledgments}
MH thanks EPSRC (UK) and the School of Mathematics and Statistics,  University of Sheffield, for a studentship supporting this work.  The work of EW is supported by the Lancaster-Manchester-Sheffield Consortium for Fundamental Physics under STFC grant ST/L000520/1.
\end{acknowledgments}

\bibliographystyle{apsrev4-1}

\end{document}